\documentclass[runningheads]{svmult}

\usepackage{makeidx}   
\usepackage{graphicx}  
\usepackage{subeqnar}  
\usepackage{multicol}  
\usepackage{cropmark} 
\usepackage{physprbb}  
\makeindex             

\newcommand{\rsiii}{\cal{R}(Si~II)}
\newcommand{\kms}{km~s$^{-1}$}
\begin{document}
\title*{Optical Spectra of Supernovae}
\toctitle{Optical Spectra of Supernovae}
\titlerunning{Spectra of Supernovae}
\author{David Branch\inst{1}
\and E. Baron\inst{1}
\and David J. Jeffery\inst{2}}
\authorrunning{Branch et al.}
%
\institute{University of Oklahoma, Norman, OK 73019, USA
\and New Mexico Institute of Mining and Technology,
Socorro, NM  87801, USA}

\maketitle              

\begin{abstract}

Supernova flux and polarization spectra bring vital information on the
geometry, physical conditions, and composition structure of the
ejected matter. For some supernovae the circumstellar matter is also
probed by the observed spectra.  Some of this information can be
inferred directly from the observed line profiles and fluxes, but
because of the Doppler broadening and severe line blending,
interpretation often involves the use of synthetic spectra.  The
emphasis in this Chapter is on recent results obtained with the help
of synthetic spectra.

\end{abstract}

\section{Introduction}

As discussed and illustrated by M.~Turatto in Chap.~3, optical spectra
serve as the basis for the classification of supernovae (SNe).  This
chapter is concerned with extracting information from optical spectra
on the geometry, physical conditions, and composition structure
(element composition versus ejection velocity) of the ejected matter,
and also the circumstellar matter for some SNe.  Several other
chapters of this volume also discuss certain aspects of SN
spectroscopy: ultraviolet spectra (Chap.~7 by N.~Panagia); the
circumstellar and nebular spectra of SN~1987A (Chap.~15 by R.~McCray);
and the hypernova SN~1998bw (Chap.~25 by K.~Iwamoto et~al.).

Some of the information carried by SN spectra can be inferred directly
from the observed line profiles and fluxes, but because SN ejection
velocities are a few percent of the speed of light, spectral features
generally are blended and interpretation often involves comparing
observed spectra with synthetic spectra calculated for model SNe.  The
spectrum calculations may be simplified and rapid, or physically
self-consistent and computationally intensive.  Similarly, both simple
parameterized physical models and numerical hydrodynamical
models are used. In this chapter we place some emphasis on recent
results obtained with the help of synthetic spectra.

In \S2 the elements of spectral line formation in SNe are discussed
and some of the synthetic spectrum codes that are in current use are
described.  In \S\S3--6 recent comparisons of synthetic spectra with
observed spectra of SNe of Types Ia, Ib, Ic, and II are surveyed.
Space limitations prevent us from referring to the numerous papers
that have been important to the development of this subject; many such
can be found in 1993 and 1997 review articles \cite{Whe93,Fil97}.  Here
most of the references are to, and all of the figures are from, papers
that have appeared since 1998. A brief discussion of the prospects for
the future appears in \S7.

\section{Line Formation and Synthetic Spectrum Codes}

At the time of the supernova explosion, physical scales are relatively
small. Shortly thereafter the ejected matter expands freely as from a
point explosion
with each matter element having a constant velocity.  All structures
in the ejecta then just scale linearly with $t$, the time since
explosion. The matter is in homologous expansion, which has several
convenient features: (1)~the radial velocity $v$ of a matter element
is a useful comoving coordinate with actual radial position of the
element given by $r=vt$; (2)~the density at any comoving point just
scales as $t^{-3}$; (3)~the photon redshift between matter elements
separated by velocity $\Delta v$, $\Delta\lambda=\lambda(\Delta v/c)$,
is time-independent; and (4)~the resonance surfaces for line emission
at a single Doppler-shifted line frequency are just planes
perpendicular to the observer's line of sight.  (Relativistic effects
introduce a slight curvature \cite{Jef93}.)

If continuous opacity in the line forming region is disregarded, the
profile of an unblended line can be calculated when the line optical
depth $\tau_\ell(v)$ and source function $S_\ell(v)$ are specified.
Because SN ejection velocities ($\sim10,000$ \kms) are much larger
than the random thermal velocities ($\sim10$ \kms), a photon remains
in resonance with an atomic transition only within a small resonance
region.  The Sobolev approximation, that the physical conditions other
than the velocity are uniform within the resonance region, usually is
a good one, and it allows the optical depth of a line to be simply
expressed in terms of the local number densities of atoms or ions in
the lower and upper levels of the transition:

$$ \tau_\ell
= {{\pi e^2} \over {m_e c}}\ f\ \lambda\ t\ n_\ell
\left(1-{{g_\ell n_u} \over {g_u n_\ell}}\right)
= 0.229\ f\ \lambda_\mu\
t_d\ n_\ell\left(1-{{g_\ell n_u} \over {g_u n_\ell}}\right),  $$

\noindent where $f$ is the oscillator strength, $\lambda_\mu$ is the
line wavelength in microns, $t_d$ is the time since explosion in days,
and $n_\ell$ and $n_u$ are the populations of the lower and upper
levels of the transition in cm$^{-3}$.  The term in brackets is the
correction for stimulated emission.  The source function is

$$ S_\ell = {2 h c \over \lambda^3}\ \left({g_u n_\ell \over g_\ell
n_u} - 1\right)^{-1}.  $$

\noindent All of the radial dependence of $\tau_\ell$ and $S_\ell$ is
in the level populations.  The specific intensity that emerges from a
resonance region is

$$ I = S_\ell \left(1 - e^{-\tau_\ell}\right). $$

Spectroscopic evolution can be divided into a photospheric phase when
the SN is optically thick in the continuum below a photospheric
velocity, and a subsequent nebular phase during which the whole SN is
optically thin in the continuum.  In the photospheric phase line
formation occurs above the photosphere and in the nebular phase
throughout the ejecta.  There is, of course, no sharp division between
the two phases.  However, spectral synthesis modeling techniques in
the photospheric and nebular limits can make use of different approximations
which are adequate for those limits.

In the remainder of this section we first discuss line formation and
synthetic spectrum codes in the photospheric and nebular phases under
the assumptions of spherical symmetry and negligible circumstellar
interaction.  We then consider the effects of circumstellar
interaction, and SN asymmetry and polarization
spectra.

\subsection{The Photospheric Phase}

The elements of spectrum formation in the photospheric phase have been
discussed and illustrated at length elsewhere \cite{Jef90}.  Here we will
only briefly summarize those elements and then discuss spectrum
synthesis codes.

During the photospheric phase a continuum radiation field is emitted
by a photosphere which can be idealized as an infinitely thin layer.
Above the photosphere the radiation interaction with continuous
opacity is small.  Line opacity on the other hand can be very large
for the strongest lines.  The large Doppler shifts spread line opacity
over a large wavelength interval increasing the effect of strong lines
compared to a static atmosphere where such strong lines saturate and
can only affect radiation in a narrow wavelength interval.  The
cumulative effect of many lines, strong and weak, can create a
quasi-continuous opacity in the Eulerian frame. This effect has been
called the ``expansion opacity'' \cite{Karp77,Pin00}; it dominates in
the ultraviolet, where it effectively pushes the photosphere out to a
larger radius than in the optical. Full, co-moving, line-blanketed
calculations automatically account for this effect \cite{Bar96a};
introducing the observer's frame expansion opacity into comoving frame
calculations is incorrect.

In the optical, which is usually the chief focus of analysis, the
spectrum is characterized by P~Cygni lines superimposed on the
photospheric continuum.  The P~Cygni profile has an emission peak near
the rest wavelength of the line and a blueshifted absorption feature.
The peak may be formed in part by true emission or by line scattering
into the line of sight of photons emitted by the photosphere.  The
emission peak would tend to be symmetrical about the line center
wavelength if not for the blueshifted absorption.  The absorption is
formed by scattering out of the line of sight of photospheric photons
emitted toward the observer.  Since this occurs in front of the
photosphere, the absorption is blueshifted.  At early times the ejecta
density is high, the photosphere is at high velocity, and the line
opacity is strong out to still higher velocities.  As expansion
proceeds, the photosphere and the region of line formation recede
deeper into the ejecta.  The P~Cygni line profile width thus decreases
with time.  The minimum of the absorption feature of weak lines tends
to form near the photospheric velocity, thus weak lines (e.g., weak
Fe~II lines) can be used to determine the photospheric velocity's time
evolution.  The recession of the photosphere exposes the inner ejecta
and permits its analysis.

In understanding P~Cygni line formation in SNe it is conceptually
useful, and often in simple parameterized radiative transfer modeling
a good physical approximation, to consider the line-photon
interaction as being pure resonance scattering; i.e., no photon
creation or destruction above the photosphere.  Then the source
function of an unblended line is just $S_\ell(v) = W(v)I_{\rm phot}$,
where $W(v)$ is the usual geometrical dilution factor \cite{Mih78} and
$I_{\rm phot}$ is the continuum specific intensity (assumed angle-independent)
radiated by the photosphere.  Given the pure resonance scattering
approximation for an unblended P~Cygni line, the emission is formed just
by scattering into the line of sight and the absorption just
by scattering out of the line of sight.

    Unfortunately, in general there is strong line blending.  A photon
that is scattered by one transition can redshift into resonance with
another, so the influence of each transition on others of longer
wavelength must be taken into account.  This multiple scattering
corresponds to the observer's line blending.  Calculations show that
in blends, absorptions trump emissions:  i.e., when resolved, the
absorption minima of blends tend to be blueshifted by their usual
amounts, but the emission peaks do not necessarily correspond to the
rest wavelengths of the lines of the blend.  Thus absorption minima
usually are more useful than emission peaks for making line
identifications during the photospheric phase.

    If one has made the resonance scattering approximation for simple
modeling, it turns out that the effect of line blending on source
functions for some lines is often effectively very small.  From a
conceptual point of view it is convenient to think of $S_\ell(v) =
W(v)I_{\rm phot}$ even for strongly blended lines.  In general the
effect of line blending on the emergent spectrum is significant
and cannot be neglected.

\begin{figure}[ht]
\begin{center}
\includegraphics[width=.9\textwidth]{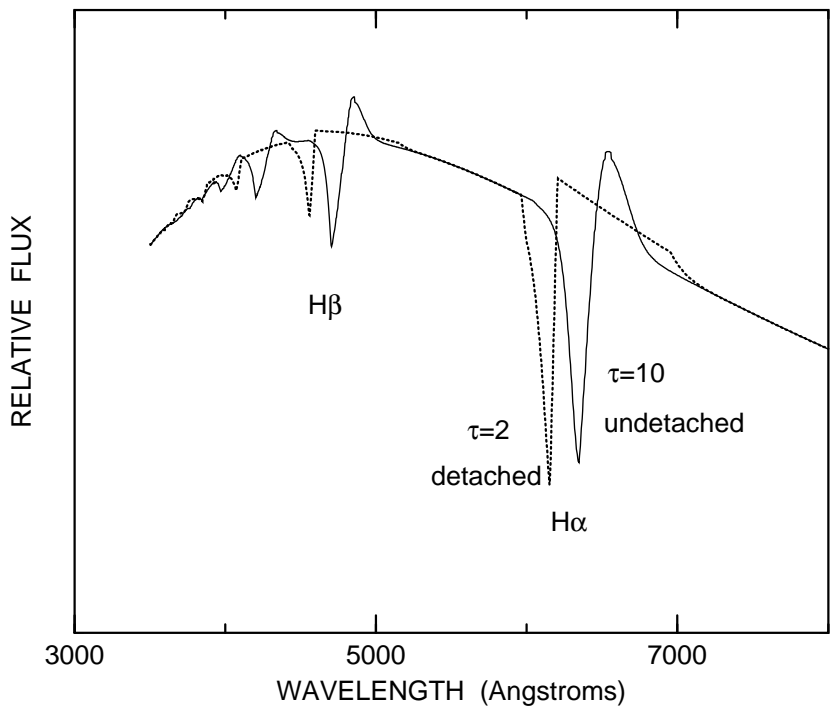}
\end{center}
\caption[]{A \texttt{SYNOW} synthetic spectrum ({\it dotted line}) that has
$v_{phot}=10,000$ \kms\ and hydrogen lines detached at 20,000 \kms\
where $\tau$(H$\alpha$)=2, is compared with a synthetic spectrum ({\it
solid line}) that has $v_{phot}=10,000$ \kms\ and undetached hydrogen
lines with $\tau$(H$\alpha$) = 10 at the photosphere.  (From
\cite{Bra02})\label{fig:detach} }
\end{figure}

    A special case of a P~Cygni line that has become of interest is a
``detached'' line: i.e, a line that has a significant optical depth
only above some detachment velocity that exceeds the velocity at the
photosphere.  A detached line consists of a flat inconspicuous
emission peak and an absorption having a sharp red edge at the
blueshift corresponding to the detachment velocity (Fig~1). We shall
later refer to the concept of detached lines in discussing particular
observed spectra.

    It is possible to extract some information from spectra by simple
direct analysis.  Recently a method for inverting an unblended P~Cygni
line profile to extract the radial dependences of the line optical
depth and the source function has been developed \cite{Kas01}.  But
the complex blending of P~Cygni lines (which often occurs), and the
need to account for effects that cannot treated by direct approaches
make analysis by synthetic spectrum modeling essential.  Rather than
discuss modeling in general we will briefly describe three synthetic
spectrum codes that are frequently used, proceeding from simple to
complex.  Example synthetic spectra from these codes will appear in
\S\S3 -- 6.

    The simplest code is the fast, parameterized \texttt{SYNOW} code.
Technical details of the current
version are in \cite{Fis00}.  The basic assumptions are: spherical symmetry;
a sharp photosphere that emits a blackbody continuous spectrum; and
the Sobolev approximation with a resonance scattering source function.
\texttt{SYNOW} does not do continuum transport, it does not solve rate
equations, and it does not calculate or assume ionization ratios.
Its main function is to take line multiple scattering into account so
that it can be used in an empirical spirit to make line
identifications and determine the velocity at the photosphere and the
velocity intervals within which the presence of each ion is detected.
For each ion that is introduced, the optical depth at the photosphere
of a reference line is a free parameter, and the optical depths of the
other lines of the ion are calculated assuming Boltzmann excitation.
Reference lines are generally chosen as the strongest line in the
optical for a particular species.
Line optical depths are assumed to decrease radially following a power
law or an exponential.  When deciding which ions to introduce, use is
made of the results of \cite{Hat99b} who presented plots of LTE Sobolev line
optical depths versus temperature for six different compositions that
might be expected to be encountered in SNe, and sample \texttt{SYNOW} optical
spectra for 45 individual ions that are candidates for producing
identifiable spectral features in SNe.  When fitting to an observed
spectrum, the important parameters are the reference line optical
depths, the velocity at the photosphere, and whatever maximum and
minimum (detachment) velocities may be imposed on each ion.

Somewhat more complex synthetic spectrum calculations are carried out
with what we will refer to as the \texttt{ML MONTE CARLO} code (ML for
Mazzali and Lucy), technical details of which are in \cite{Maz00a} and
references therein.  This code also assumes a sharp photosphere, makes
the Sobolev approximation, and does not solve rate equations.  The
main differences from \texttt{SYNOW} are: (1) an approximate radiative
equilibrium temperature distribution and internally consistent
ionization ratios are calculated for an assumed composition structure;
(2) electron scattering in the atmosphere is taken into account; and
(3) the original assumption of resonance scattering has been replaced
by an approximate treatment of photon branching, to allow an
absorption in one transition to be followed by emission in another
\cite{Luc99}.  Several circumstances in which the effects of photon
branching are significant are discussed in \cite{Maz00a}.

Very detailed calculations are made with the multi-purpose synthetic
spectrum and model atmosphere code called \texttt{PHOENIX}.  Technical
details of \texttt{PHOENIX} are in \cite{Hau99} (see also
\cite{Bar96b}).  The basic assumptions are spherical symmetry and time
independence. The Sobolev approximation is dispensed with, and
calculations are carried out in the comoving frame with all special
relativistic effects.  The aim of \texttt{PHOENIX} is to take all of
the relevant physics into account as fully as possible.
\texttt{PHOENIX} treats continuum transport, solves the NLTE rate
equations for a large number of ions and a very large number of atomic
levels with gamma ray deposition and nonthermal excitation taken into
account, and determines a self-consistent radiative equilibrium
temperature structure.  The lower boundary condition can be chosen to
be either diffusive or nebular.

    One may ask why simpler codes are needed when \texttt{PHOENIX}-like codes
can be implemented.  \texttt{PHOENIX}--like codes demand many hours of
computation and are not optimal for explorations of parameter space
or for gaining insight into how simple features of modeling affect
outcomes.  The simpler codes are tools for rapid exploration and
gaining insight.  Such codes, however, cannot give definitive
determinations because of their simplified physics.  \texttt{PHOENIX}--like
calculations must be the basis for ultimate decisions about the
viability of SN models.

\subsection{The Nebular Phase}

Our discussion of line formation in the nebular phase will be brief
since the subject is discussed in Chap.~15, in the context
of SN~1987A. (See also \cite{Fra94}.)

In the absence of circumstellar interaction, the nebular phase is
powered by radioactive decay, primarily $^{56}$Co (77.23~day
half-life), the daughter of explosion-synthesized $^{56}$Ni (6.075~day
half-life).  At very late times longer lived radioactive species such
as $^{57}$Co and $^{44}$Ti become important.

The decay gamma~rays deposit their energy by Compton scattering off
electrons to produce nonthermal fast electrons whose energy quickly
goes into atomic ionization and excitation, and Coulomb heating of the
thermal electrons \cite{Swa95}.  Particularly for low mass SNe
(Types~Ia,b,c) the unscattered escape of gamma~rays (which increases
as the ejecta thins) causes the kinetic energy of positrons from the
$^{56}$Co decays to become an important energy source: about $3\,$\%
of the $^{56}$Co decay energy is in the form of positron kinetic
energy.  The positrons are much more strongly trapped than the gamma
rays---but not completely trapped \cite{Mil01}. The positron
kinetic energy goes into creating fast electrons leading to the same
deposition processes as the gamma~rays.  Optical emission lines are
formed by recombination, collisional excitation, and fluorescence.
P~Cygni profiles of permitted lines that form by scattering in the
outer layers may be superimposed on the emission line spectrum.

Because of low continuous opacity and increasingly low line opacities,
photons emitted by true emission processes are increasingly unlikely
to be scattered again.  Thus non-local radiative transfer effects in
the nebular phase tend to become negligible.  (Local trapping in an
optical thick line can be treated easily using a Sobolev escape
probability.)  Thus (absent circumstellar interaction) it is somewhat
more feasible than in the photospheric phase to extract information
without making synthetic spectrum calculations.  This was especially
true of SN~1987A, which because of its proximity was well observed in
the optical and the infrared long into the nebular phase when the
characteristic velocity of the line forming region was only $\sim3000$
\kms\ and blending was not too severe (see Chap.~15).  Most other
non-circumstellar interacting SNe are observed only for a year or so
after explosion and line blending is more of a problem.

Without a photosphere, there is no continuum flux to be scattered
out of the line of sight even if there are optically thick lines.
Hence the characteristic line profile of the nebular phase is not a P~Cygni
profile, but an emission line that peaks near the rest wavelength and
for several reasons may have an extended red wing \cite{Fra89,Chu00a}.
If the line has significant optical depth down to low velocity, the
emission line has a rounded peak.  If instead the line forms in a
shell, the emission has a boxy (flat top) shape with the half width of
the box corresponding to the minimum velocity of the shell.

As for the photospheric phase, nebular-phase synthetic spectrum
calculations of various levels of complexity prove to be useful.
Relatively simple parameterized approaches that can be used as
diagnostic tools are described by \cite{Bow97,Fra01}.  A one-zone code
that was developed in its original form by \cite{Rui92} and will be
referred to here as the \texttt{RL NEBULAR} code (RL for Ruiz-Lapuente
and Lucy) is frequently used (e.g. \cite{Maz01a,Maz97,Rui95}).  The
input parameters are the time since explosion and the mass,
composition, and outer velocity of the nebula.  Heating is calculated
from the $^{56}$Co decay rate, and cooling is by radiative emission.
The electron density, temperature, NLTE level populations and emission
line fluxes are determined simultaneously by enforcing statistical and
thermal equilibrium.  Codes for calculating synthetic spectra of
hydrodynamical models are described by \cite{Hou96} and \cite{Liu98}
and references therein.  These multi-zone codes calculate the
nonthermal heating, ionization equilibrium, and the NLTE level
populations for a large number of atomic levels.  Synthetic spectra of
the CO molecule have also been calculated \cite{Fas01,Ger00a,Spy01}.
For a description of additional nebular phase codes that have been
used for SN~1987A, see Chap.~15.

The photospheric phase forbids a direct view into the inner regions of
the ejecta.  The nebular phase permits this direct view, but, on the
other hand, the outer layers tend toward invisibility.  The
photospheric phase demands a powerful radiative transfer technique.
The advanced nebular phase requires almost none at all, at least in
the optical; the ionizing photons in the UV may need a detailed
radiative transfer treatment and, of course, the gamma ray and
positron energy depositions need a non-local transfer treatment
probably best done by Monte Carlo.  Both phases require NLTE
treatments and extensive atomic data.  The two phases offer
complementary insights and challenges for analysis.

\subsection{Circumstellar Interaction}

The physics of the radiative and hydrodynamical interactions between
SNe and their circumstellar matter (circumstellar interaction: CSI)
are discussed in Chap.~14 by R.A.~Chevalier \& C.~Fransson.  No
convincing detection of SN~Ia CSI has yet been made in any wavelength
band.  CSI also does not seem to affect the optical spectra of typical
SNe~Ib, SNe~Ic (but for an exception see \cite{Mat00a}), and many SNe~II.
Some SNe~II and perhaps some hyperenergetic SNe~Ic are affected, and
in extreme cases dominated, by CSI effects.

In the idealized case, the wind of the SN progenitor star has a
constant velocity and a constant mass loss rate, and therefore an
$r^{-2}$ density distribution.  After the explosion, the circumstellar
matter can be heated and accelerated by the photons emitted by the SN.
Subsequent hydrodynamic interaction between high velocity SN ejecta
and low velocity circumstellar matter generates photon emission that
further affects the velocity distribution and the ionization and
excitation of both the circumstellar matter and the SN ejecta.  These
interactions can produce a rich array of effects on the optical
spectra.  During the photospheric phase narrow emission, absorption,
and P~Cygni lines formed in low velocity circumstellar matter may be
superimposed on the SN spectrum \cite{Ben98,Ben99,Sal98,Sol98a}. The SN
resonance scattering features may be ``muted'' by the external
illumination of the SN line forming region \cite{Bra00}.  In extreme cases
the circumstellar matter may be optically thick in the continuum so
that the spectrum consists only of circumstellar features, perhaps
including broad wings produced by multiple scattering off
circumstellar electrons \cite{Chu01}.  During the nebular phase CSI can
power boxy emission from high velocity SN matter \cite{Fra01} and at very
late times, even decades, CSI can allow the detection of SNe that
would otherwise be unobservably faint \cite{Fes99}.

Because the simple $r=vt$ velocity law does not apply to circumstellar
matter, and blending of circumstellar lines generally is not as severe
as it is for SN lines (and spherical symmetry often is not an adequate
approximation for circumstellar matter), few calculations of synthetic
spectra for CSI have been carried out so far.  Line profiles for
various velocity and density distributions have been calculated
\cite{Fra84,Chu02} and some \texttt{PHOENIX} calculations for a
constant velocity circumstellar shell have been carried out
\cite{Len01a}.  Most of the analysis of optical CSI features has
involved extracting information on velocities directly from individual
line profiles, and on physical conditions and abundances from emission
line fluxes \cite{Che94,Chu02,Fas01,Mat00b}.

\subsection{Asymmetry and Polarization Spectra}

     It has been the heuristic and hopeful assumption that SNe are
essentially spherically symmetric.  But from SN polarimetry,
especially spectropolarimetry, it is now clear that all types of SNe
other than SNe~Ia usually do exhibit some kind of significant
asymmetry (e.g, \cite{How01,Leo00b,Wan01a,Whe00}).

     The polarization of SN flux in most cases arises from their
atmospheres where continuous opacity is dominated by electron
scattering in the optical and near infrared.  Electron scattering is
polarizing with polarization (for incident unpolarized radiation)
varying from $100\,$\% for $90^{\circ}$ scattering to $0\,$\% for
forward and backward scattering according to the Rayleigh phase matrix
\cite{Cha60}.  The polarization position angle is perpendicular to the
scattering plane defined by incident and scattered light.  The
emergent flux from a deep electron scattering atmosphere (either
plane-parallel or geometrically extended) tends to exhibit a position
angle aligned perpendicular to the normal to the surface and
polarization increasing monotonically from $0\,$\% at an emission
angle of $0^{\circ}$ (to the normal) to a maximum at an emission angle
of $90^{\circ}$.  For the plane-parallel case the maximum is only
$11.7\,$\% \cite{Cha60}, but for a highly extended atmosphere it can
approach $100\,$\% \cite{Cas71} from the limb, if it can be resolved.  SN
atmospheres fall into the extended class, and so beams from their limb
should be highly polarized though mostly not at $100\,$\%.  Beams from
the photodisk region (the projection of the photosphere on the sky)
should be much less polarized.  If SNe were spherically symmetric, the
integrated flux (which is all that can be observed from extragalactic
SNe) would have zero polarization since all polarization alignments
would contribute equally and thus cancel producing zero net
polarization.
Because of cancellation, the predicted net continuum polarizations
are low (a few percent:  see below) even for quite strong asymmetries.

Line scattering is usually depolarizing \cite{Jef89}.  The simplest line
polarization profile is an inverted P~Cygni profile that is common
although not universal in the SN~1987A data \cite{Jef91} and is seen in
some of the later spectropolarimetry \cite{Wan01a}.  In this profile a
polarization maximum is associated with the P~Cygni absorption feature
of the flux spectrum and a polarization minimum is associated with the
P~Cygni flux emission feature.  The inverted P~Cygni profile has a
natural explanation.  The polarized flux at any wavelength tends to
come from $90^{\circ}$ electron scattering.  At the flux absorption,
unpolarized flux directly from the photosphere is scattered out of the
line of sight by a line: thus, the electron polarized flux is less
diluted and a polarization maximum arises.  At the flux emission, the
line scatters into the line of sight extra unpolarized flux diluting
the polarization due to the electron scattered flux.  Almost any shape
asymmetry should tend to give the inverted polarization P~Cygni
profile.  Thus, the inverted P~Cygni profile is somewhat limited as a
direct analysis tool for asymmetry.

Although one must be cautious in generalizing because of the paucity
and inhomogeneity of published data, it seems that most
core-collapse SNe that have been observed well
enough to detect polarization of order a few tenths of percent (of
order 40 to date) have shown at least that much continuum
polarization.  Some of this can be interstellar polarization (ISP) due
to intervening dust, but intrinsic polarization seems to be present in
many cases.  The intrinsic polarization reveals itself through time
dependence and, in spectropolarimetry, by line polarization features or
by deviation from  the ISP wavelength dependence \cite{Ser73}.  Unfortunately,
correcting for parent--galaxy ISP can only be done from an analysis of
the SN data itself because there is no other bright nearby point
source.  Even the Galactic ISP correction may often have to be
determined using
the SN data since ISP can vary rapidly with position and depth and
there may be no suitable Galactic star near the angular position of
the SN.  Techniques for correcting for ISP have been developed
\cite{How01,Jef91,Leo00a}.

Fig.~2 shows a sample of spectropolarimetry from a range of SN types
overlapping the corresponding relative flux spectra.  The polarization
has been corrected for ISP, but in most cases only with the intention
of highlighting the line polarization features: the overall
polarization level is not trustworthy.  Some of the line polarization
features are consistent with the inverted P~Cygni profile.  However,
line polarization features are badly distorted by ISP, and thus
reliable ISP corrections are needed to determine the true line
polarization profiles as well as the true continuum polarization.

\begin{figure}[ht]
\begin{center}
\includegraphics[width=.7\textwidth]{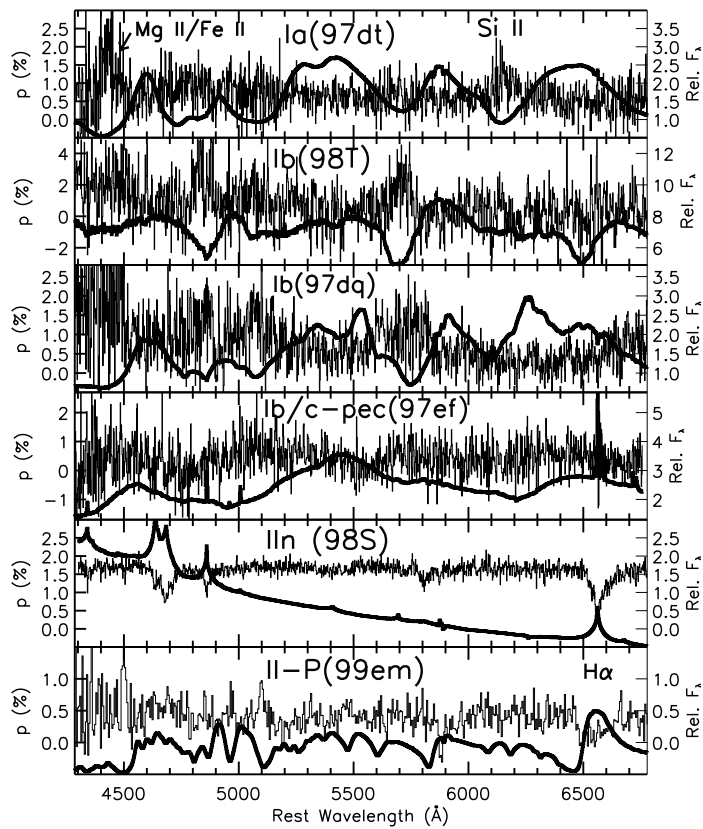}
\end{center}
\caption[]{Spectropolarimetry ({\it thin lines}) of various SN
types, with the corresponding flux spectra ({\it thick lines}), all
obtained within two months of explosion.  Figure courtesy of
D.C.~Leonard}
\label{fig:polar}
\end{figure}

    For core-collapse SNe the intrinsic continuum polarization varies
from probably $\sim 0.0\,$\% up to perhaps $\sim 4\,$\%
\cite{Leo01a,Wan01a}: a representative value would be $1\,$\%.  There
may be a class of core-collapse SNe (those with massive hydrogen
envelopes) that have nearly zero polarization at early times
\cite{Leo01a}, and so are highly spherical.  Apparent trends are that
polarization increases with decreasing envelope mass (i.e., SNe~II-P
least polarized, SNe~Ic most), and that polarization increases with
time until the electron optical depth becomes small
\cite{Jef91,Leo01c,Wan01a}.  The position angle of polarization tends
to be nearly constant in time and wavelength
\cite{Jef91,Leo01c,Wan01a}, which is naturally accounted for by
axisymmetry in the ejecta.

    For SNe~Ia most observations (of order 15 to date) have only been
able to place an upper limit of about $0.3\,$\% continuum polarization
(not distinguishing intrinsic from ISP in most cases) \cite{Whe00}.
One normal event, SN~1996X, has a marginal detection of intrinsic
polarization of $\sim 0.3\,$\% \cite{Wan97}.  Spectropolarimetry of
SN~1997dt also may show the signatures of intrinsic polarization
(Fig.~2).  So far it seems that normal SNe~Ia are not very polarized
and are quite spherically symmetric: this is consistent with the
observational homogeneity that makes them such useful tools for
cosmology.  On the other hand, the subluminous SN~1991bg-like event
SN~1999by had intrinsic continuum polarization reaching up to perhaps
$0.8\,$\% \cite{How01}.  Based on a sample of one, subluminous SNe~Ia
may be significantly asymmetric.

     Various models of asymmetry have been presented to interpret the
polarization data.  The conventional model is axisymmetric ellipsoidal
asymmetry, either prolate or oblate \cite{Hof91,Jef89,Mcc84,Sha82}.
Even when viewed equator-on, considerable ellipsoidal asymmetry is
required to account for SN-like continuum polarization.  For example,
Monte Carlo calculations suggest that iso-density contour axis ratios
of order 1.2:1 and 2.5:1 are needed for polarizations of $1\,$\% and
$3\,$\%, respectively \cite{Hof96,Wan01a}.  Late-time imaging of
SN~1987A shows asymmetries suggesting axis ratios of 2:1
\cite{Hof01,Wan01b}.  Thus large organized asymmetries something like
the ellipsoidal models may account for most SN polarization.  Other
suggested models of polarizing large--scale asymmetry are asymmetric
clumpy ionization \cite{Chu92}, or scattering of SN light off a nearby
dust cloud \cite{Wan96} or off bipolar jets \cite{Jef01}.

     Spectropolarimetry plays an extremely important role in
polarimetric observations for two reasons: (1) it is very useful for
establishing intrinsic polarization and correcting for ISP, and (2) it
is probably vital for understanding the SN asymmetry.  Currently the
most advanced synthetic polarization spectrum analyses are done with a
3-dimensional (3D) NLTE code that uses Monte Carlo radiative transfer
in the outer region where emergent polarization is formed
\cite{Hof01,How01}.  The NLTE component of these calculations is
necessarily more limited than in 1D calculations since 3D radiative
transfer calculations are computationally very demanding.  The
ellipsoidal asymmetry used in the calculations of \cite{How01} is
parameterized and then fitted to the flux and polarization
observations.  Explosion model asymmetries can also be used
\cite{Hof01}.  Because of the large parameter space to explore in
investigating SN asymmetry, less elaborate codes than that of
\cite{Hof01} and \cite{How01} will also continue to be useful.  Such
codes will mostly rely on Monte Carlo radiative transfer because of
its flexibility and robustness.

In addition to the shape asymmetries discussed above, the ejected
matter also may be ``clumped'' (see Chap.~15).  Both observation and
theory indicate that macroscopic (not microscopic) mixing is
ubiquitous in SNe.  Clumping is difficult to detect via polarization
(although see \cite{Chu92}), but it can affect both the photospheric
\cite{Chu96,Fas99,Fas98} and nebular
\cite{Bow97,Chu94,Fes99,Li93,Mat00b,Maz01b} flux spectra.  A version
of \texttt{SYNOW} for calculating flux spectra with clumps
(\texttt{CLUMPYSYN}) has been developed for analysis of the
photospheric phase \cite{Tho02}.

Circumstellar matter may, and probably often does, have both global
asymmetry and clumping \cite{Chu94,Fas01,Fra01,Leo01a}.

\section{SNe Ia}

SNe~Ia are thought to be thermonuclear disruptions of accreting or
merging white dwarfs.  They can be separated on the basis of their
photospheric spectra into those that are normal and those that are
peculiar.  Normals such as the recently well observed SNe~1996X
\cite{Sal01} and 1998bu \cite{Her00,Jha99} have P~Cygni features due
to Si~II, Ca~II, S~II, O~I, and Mg~II around the time of maximum
brightness, and develop strong Fe~II features soon after maximum.
Peculiar, weak SN~1991bg--like events such as SNe~1998de \cite{Mod01},
1997cn \cite{Tur98b}, and 1999by \cite{Gar01,How01,Vin01} have, in
addition, conspicuous low excitation Ti~II features. Peculiar powerful
SN~1991T--like events such as SNe~1997br \cite{Li99} and 2000cx
\cite{Li01} have conspicuous high excitation Fe~III features at
maximum and the usual SN~Ia features develop later.  Events such as
SN~1999aa \cite{Li00} appear to link the 1991T--like events to the
normals \cite{Bra01a}.  Nebular spectra of SNe~Ia are dominated by
collisionally excited forbidden emission lines of [Fe~III], [Fe~II],
and [Co~III].  SN~Ia ejecta basically consist of a core of iron peak
isotopes (initially mostly $^{56}$Ni) surrounded by a layer of
intermediate--mass elements such as silicon, sulfur, and calcium.
Inferring the composition of the outermost layers is difficult
\cite{Fis99,Hat99a,Maz01a,Maz98b} because the lines are broad and line
formation occurs in these layers for only a short time after
explosion.

Current viable hydrodynamical models for SNe~Ia include deflagrations
and delayed detonations (see Chap.~12 by S.E.~Woosley).  The
composition structure of the classic deflagration model W7
\cite{Nom84,Thi86} can account reasonably well for most of the
photospheric and nebular spectral features of normal SNe~Ia (Figs. 3
and 4).  On the other hand, near infrared spectra appear to contain
valuable diagnostics of the composition structure
\cite{Bow97,Her00,Mei96} and may provide evidence in favor of delayed
detonation models \cite{Whe98}.

\begin{figure}[ht]
\begin{center}
\includegraphics[width=.8\textwidth]{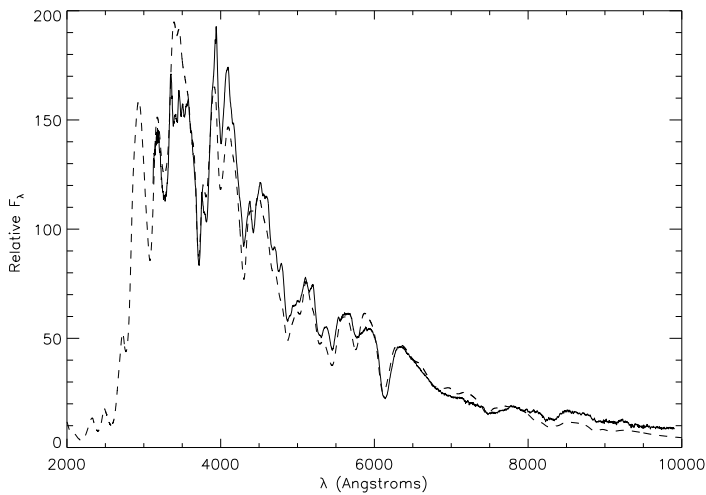}
\end{center}
\caption{The spectrum of the Type~Ia {SN~1994D} three days before
maximum brightness ({\it solid line}) and a \texttt{PHOENIX} synthetic
spectrum for the deflagration model W7 17 days after explosion ({\it
dashed line}) (From \cite{Len01c}).\label{fig:d17}}
\end{figure}

\begin{figure}[ht] \begin{center}
\includegraphics[width=.7\textwidth]{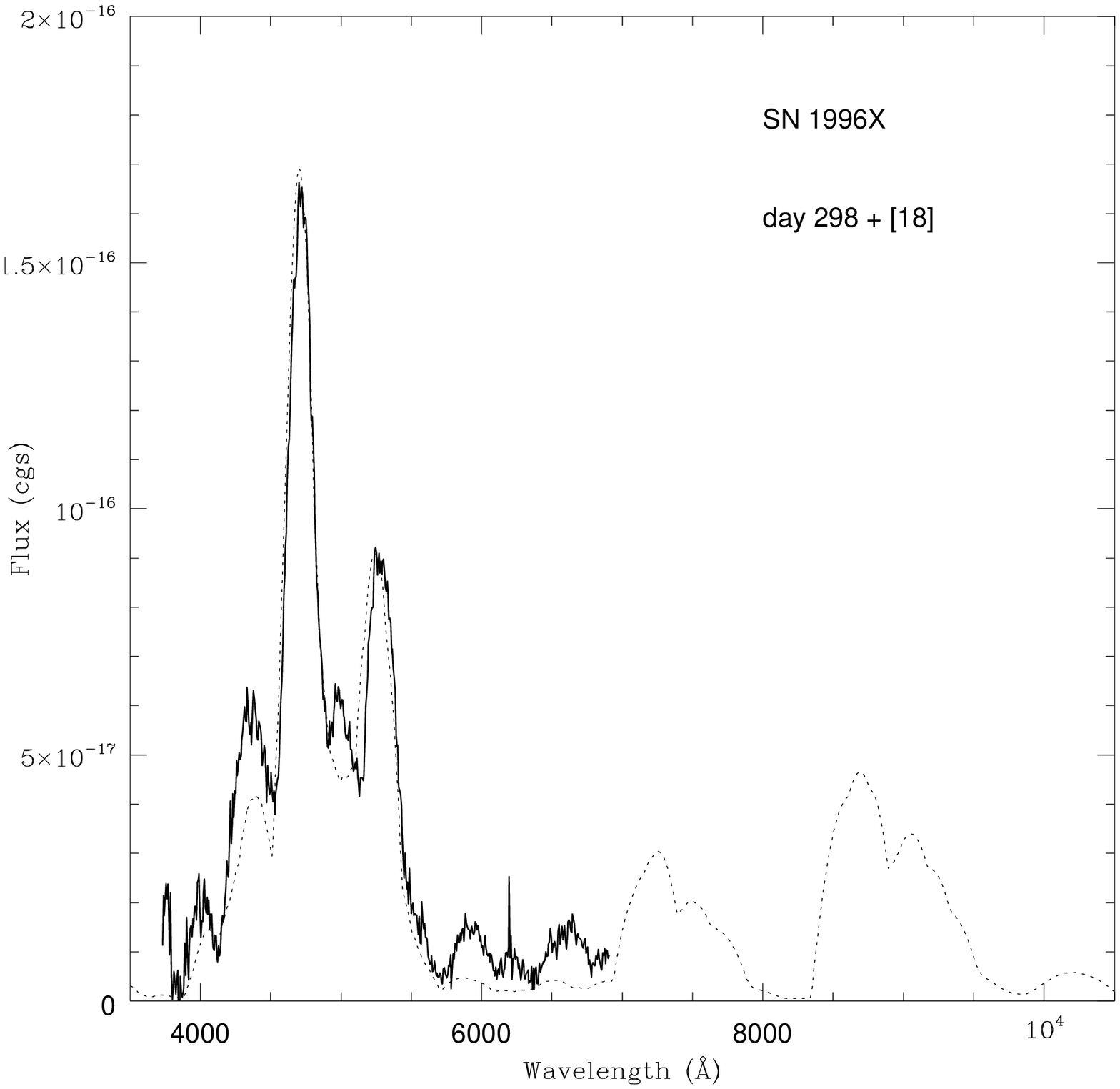} \end{center}
\caption{The spectrum of the Type~Ia SN~1996X 298 days after maximum
brightness ({\it solid line}) is compared with a synthetic spectrum
calculated with the \texttt{RL NEBULAR} code ({\it dotted line}) (From
\cite{Sal01})\label{fig:96x}}
\end{figure}

SNe~Ia can be arranged in a spectroscopic sequence ranging from the
powerful SN~1991T--like events through the normals to the weak
SN~1991bg--like events.  The sequence usually is based on the \rsiii\
parameter \cite{Nug95} -- the ratio of the depth of an absorption
feature near 5800~\AA\ to that of the deep Si~II absorption feature
near 6100~\AA.  The ratio increases with decreasing temperature owing
to the increasing contribution of numerous weak Ti~II lines to the
5800~\AA\ absorption \cite{Gar01} (Fig.~5).  The velocity at the outer
boundary of the iron peak core, as inferred from nebular spectra, also
varies along the spectral sequence \cite{Maz98a}.  The principle
physical variable along the sequence is thought to be the mass of
ejected $^{56}$Ni in a constant total ejected mass which is near the
Chandrasekhar mass.

\begin{figure}[ht]
\begin{center}
\includegraphics[width=.8\textwidth, angle=-90]{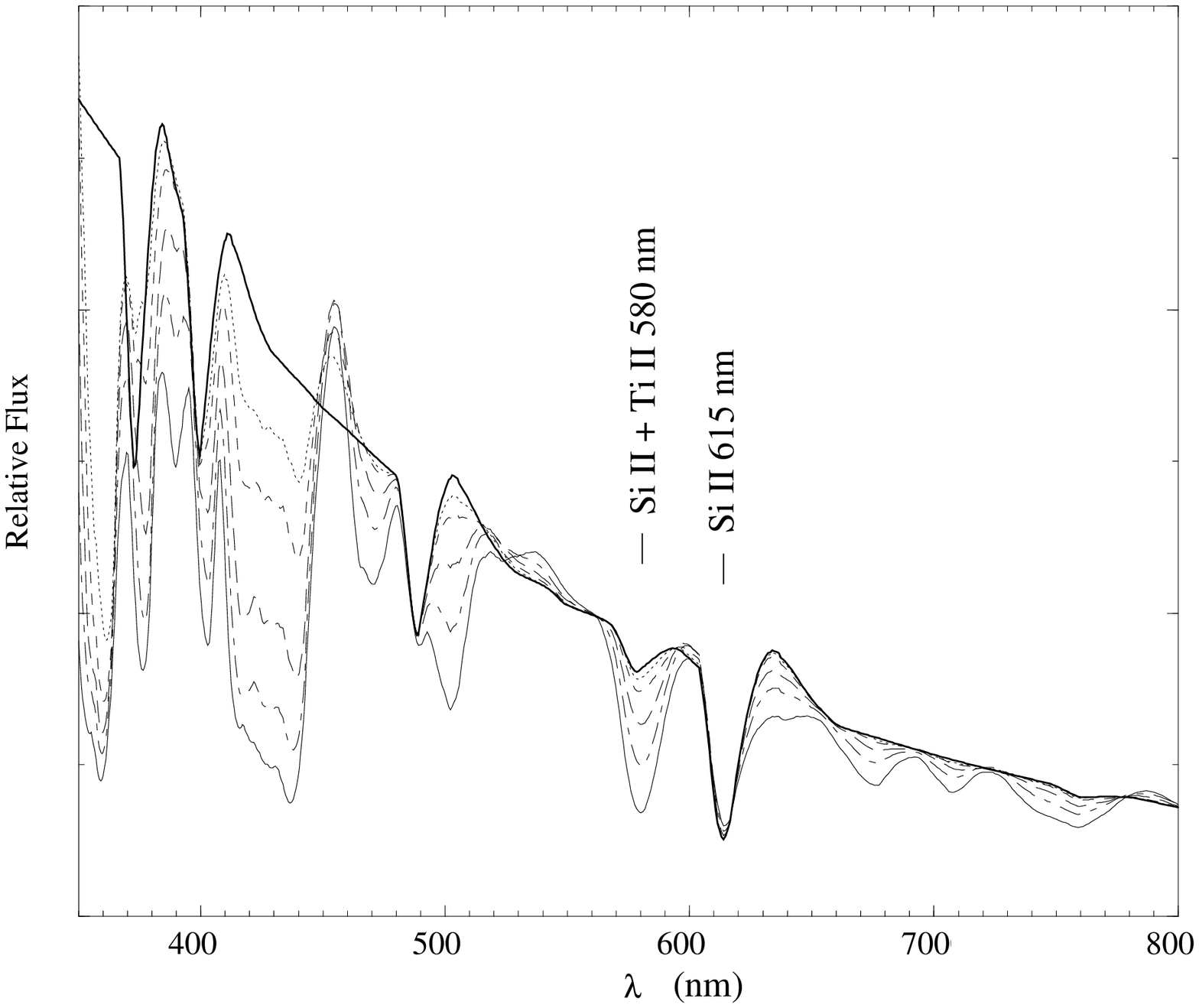}
\end{center}
\caption{\texttt{SYNOW} synthetic spectra with only Si~II lines ({\it
heavy line}) and also with Ti~II lines of various strengths ({\it thin
lines}).  The 580~nm absorption is strongly affected by Ti~II lines
(From \cite{Gar01}).\label{fig:garn}}

\end{figure}

The lack of a clear correlation between \rsiii\ and the blueshift of
the 6100~\AA\ feature shows that the diversity among SN~Ia spectra
actually is at least two dimensional \cite{Hat00}.  Blueshift
differences among events that have similar values of \rsiii\ appear to
be caused by differences in the amount of mass that is ejected at high
velocity, $\sim15,000 - 20,000$ \kms\ \cite{Len01b} (Fig~6).  One
possibility is that the high velocity events are delayed detonations
while the others are deflagrations.

\begin{figure}
\begin{center}
\includegraphics[width=0.7\textwidth]{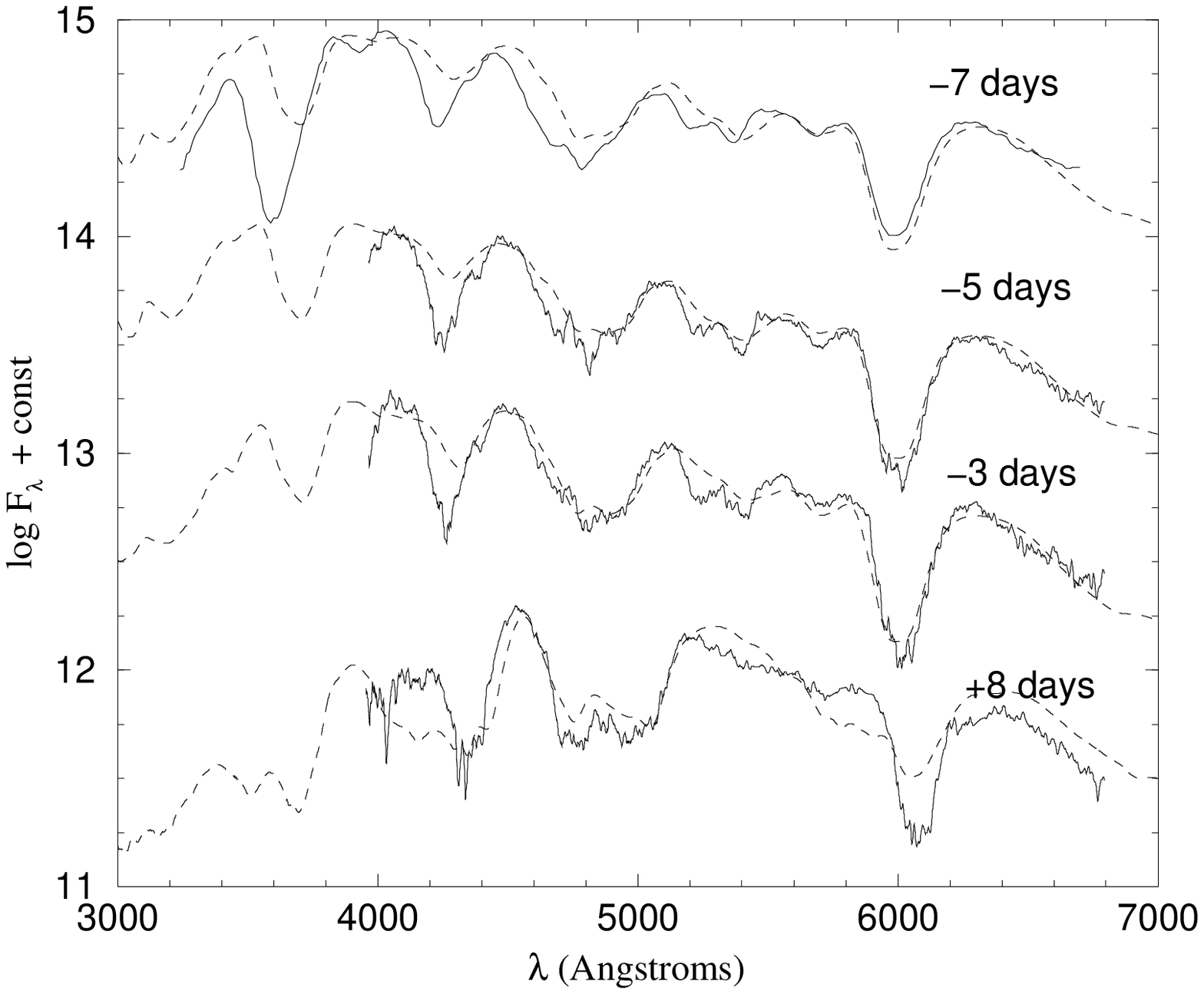}
\end{center}
\caption{\texttt{PHOENIX} synthetic spectra for the delayed detonation
model CS15DD3 ({\it dashed lines}) and observed spectra of the
high-velocity Type~Ia SN~1984A~({\it solid lines}). (From
\cite{Len01b})\label{fig:cdd3}}
\end{figure}

\begin{figure}[ht] \begin{center}
\includegraphics[width=1.0\textwidth, angle=-90]{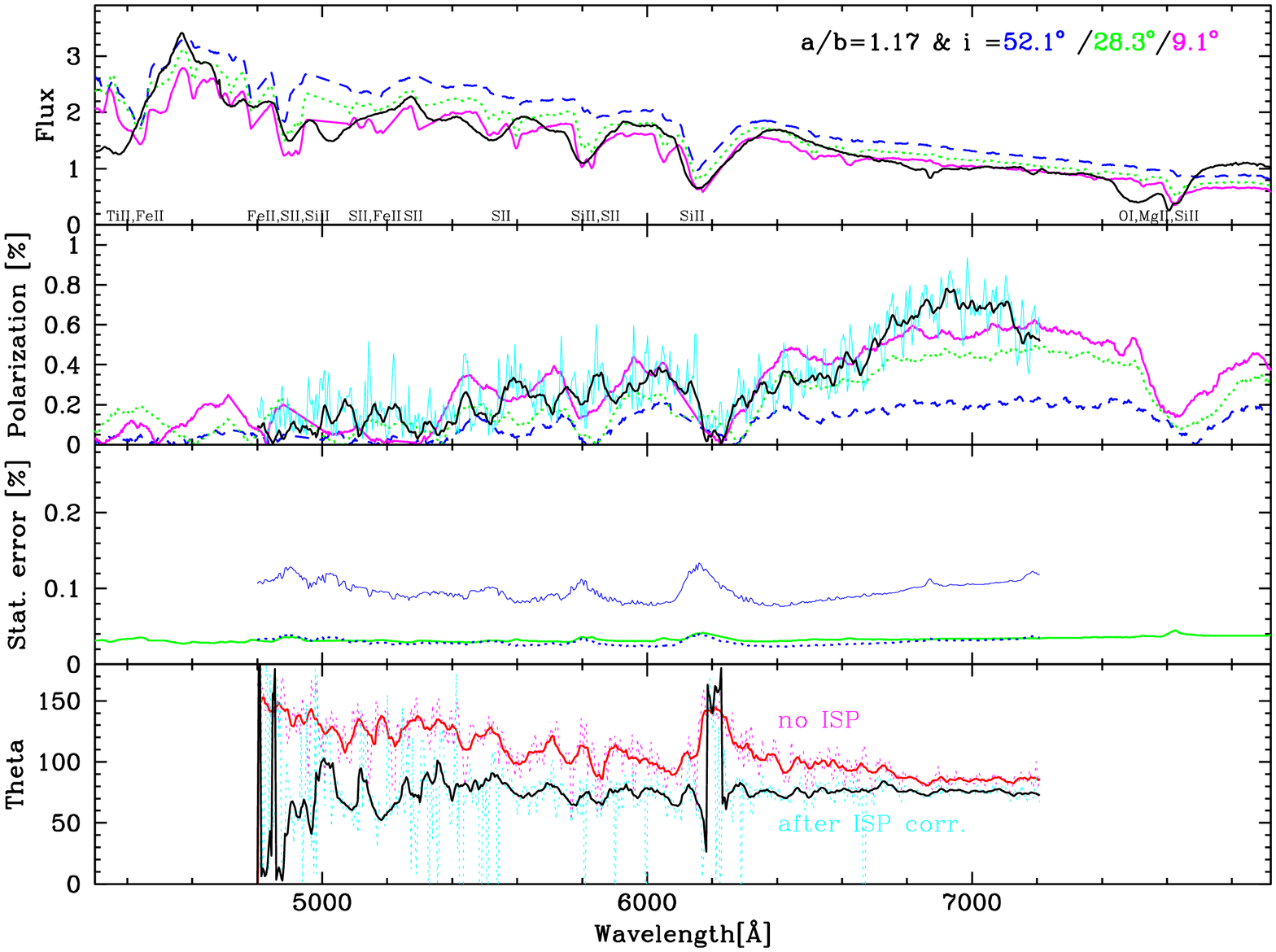}
\end{center} \caption{Flux spectra ({\it top panel, black line}) and
raw and smoothed polarization spectra ({\it second panel, light blue
and black lines}) of the peculiar weak Type~Ia SN~1999by near maximum
brightness, are compared with synthetic spectra computed for an oblate
ellipsoidal model with axis ratio 1.17 and three inclinations of the
symmetry axis from the line of sight.  (From \cite{How01}, where
further explanation of this figure can be found)\label{fig:howell}}
\end{figure}

The influence of metallicity on SN~Ia spectra is of special interest
in connection with the use of high redshift SNe~Ia as distance
indicators for cosmology.  The main effects are in the ultraviolet but
there also are some mild effects in the optical \cite{Hof98,Len00}.

Deflagrations are expected to produce macroscopic mixing and clumping
in the ejecta.  The first application of the \texttt{CLUMPYSYN} code
has shown that the uniformity of the observed depth of the 6100~\AA\
Si~II absorption in SNe~Ia near maximum brightness provides limits on the
sizes and numbers of clumps \cite{Tho02}.

Although the polarization of most SNe~Ia is low, the significant
polarization of the SN~1991bg--like event SN~1999by (Fig.~7) raises
the question of whether the production of weak SNe~Ia may involve
rapid white dwarf rotation or merging \cite{How01}.

\section{SNe Ib}

SNe~Ib are thought to result from core collapse in massive stars that
have lost all or almost all of their hydrogen envelopes.  The
photospheric spectra contain the usual low excitation SN features such
as Ca~II and Fe~II together with strong He~I lines that are
nonthermally excited by the decay products of $^{56}$Ni and $^{56}$Co
\cite{Har87,Luc91,Swa93}.  Recent work on photospheric spectrum
calculations includes a comparison of one NLTE synthetic spectrum of a
hydrodynamic model with a spectrum of SN~1984L \cite{Woo97}, a
\texttt{SYNOW} study of line identifications in SN~1999dn
\cite{Den00}, and a comparative \texttt{SYNOW} study of the spectra of
a dozen SNe~Ib \cite{Bra02}.

\begin{figure}[ht]
\begin{center}
\includegraphics[width=.8\textwidth]{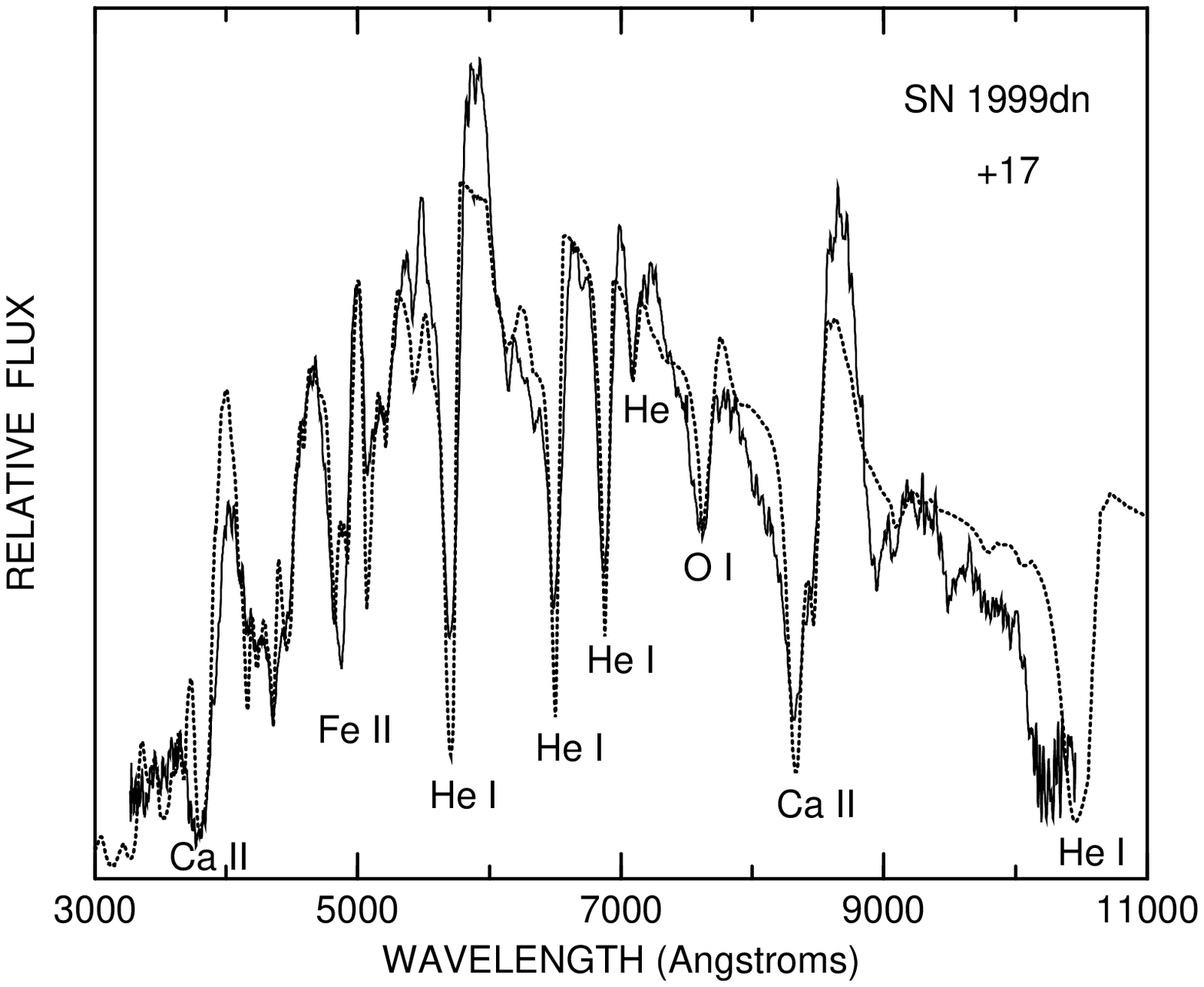}
\end{center}
\caption[]{The spectrum of the Type~Ib SN~1999dn 17~days after maximum
brightness ({\it solid line}) is compared with a \texttt{SYNOW}
synthetic spectrum ({\it dotted line}) that has $v_{phot}=6000$ \kms.
(From \cite{Bra02})}
\label{fig:99dn}
\end{figure}

\begin{figure}[ht]
\begin{center}
\includegraphics[width=.7\textwidth]{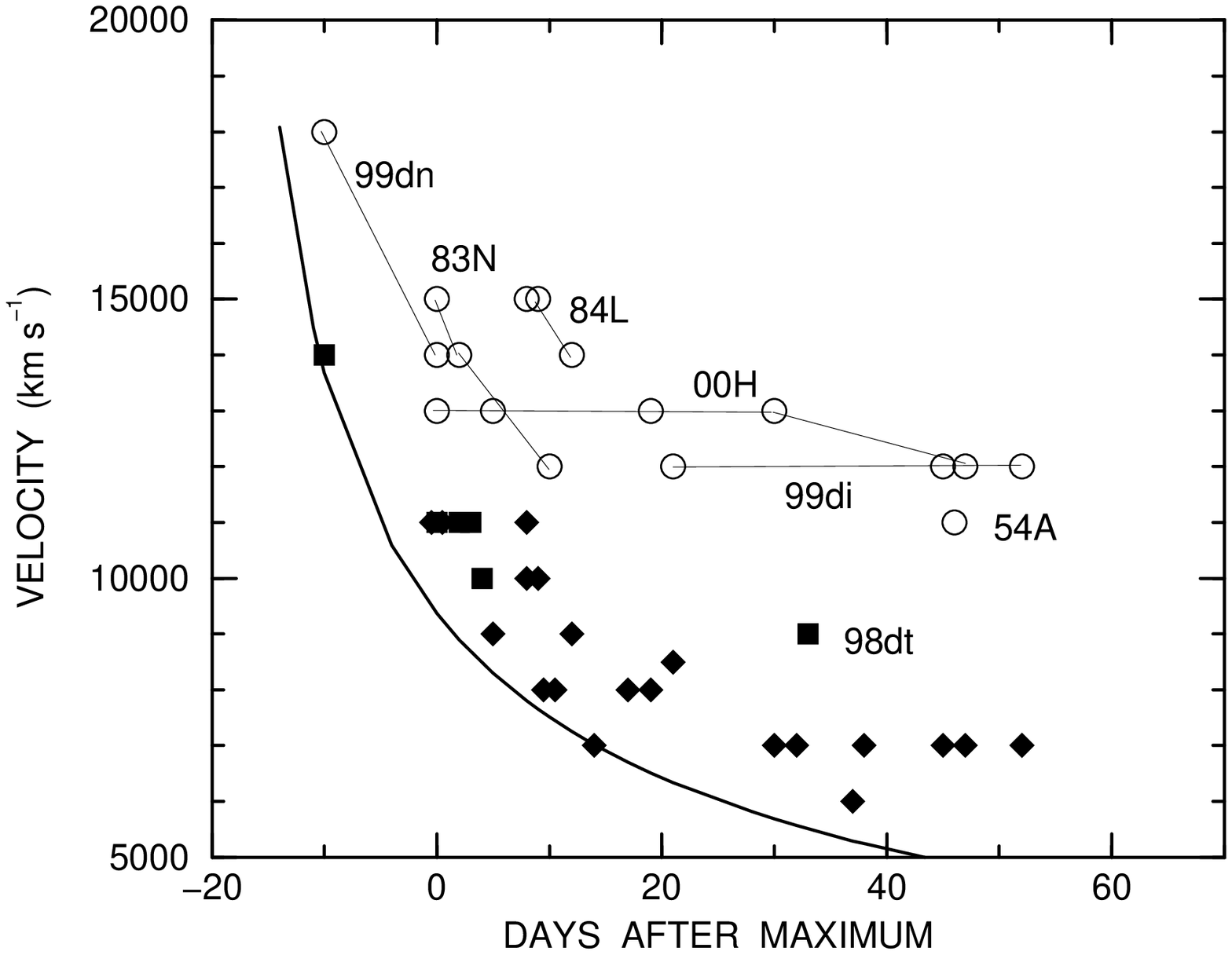}
\end{center}
\caption[]{For a sample of 12~SNe~Ib, the minimum velocity of the He~I
lines ({\it filled squares} when undetached, {\it filled diamonds}
when detached) and the minimum velocity of the hydrogen lines ({\it
open circles}, always detached) are plotted against time after maximum
brightness.  The {\it heavy curve} is a power law fit to the velocity at
the photosphere as determined by Fe~II lines. (From
\cite{Bra02})\label{fig:Ib} }
\end{figure}

\texttt{SYNOW} spectra can match the observed spectra rather well
(Fig.~8).  Consistent with the observational homogeneity of SNe~Ib
found by \cite{Mat01}, the sample studied by \cite{Bra02} obeys a
surprisingly tight relation between the velocity at the photosphere as
inferred from Fe~II lines and the time relative to maximum brightness.  The
masses and kinetic energies of the events in the sample appear to be
similar, and not much room is left for any influence of departures
from spherical symmetry on the velocity at the photosphere.  After
maximum brightness the He~I lines usually are detached, but the minimum
velocity of the ejected helium is at least as low as 7000~\kms\
(Fig.~9).  The spectra of SNe~2000H, 1999di, and 1954A contain
detached hydrogen absorption features forming at $\sim12,000$ \kms\
(Fig.~9), and hydrogen appears to be present in SNe~Ib in general,
although in most events it becomes too weak to identify soon after
maximum brightness.  The hydrogen line optical depths used to fit the
spectra of SNe~2000H, 1999di, and 1954A are not high, so only a mild
reduction would be required to make these events look like typical
SNe~Ib.  Similarly, the He~I line optical depths in typical SNe~Ib are
not very high, so a moderate reduction would make them look like
SNe~Ic.

The number of SNe~Ib for which good spectral coverage is available is
still small.  More events need to be observed to explore the extent of
the spectral homogeneity and to determine whether there is a continuum
of hydrogen line strengths.  Also needed are detailed NLTE
calculations for hydrodynamical SN~Ib models having radially
stratified composition structures, to determine the hydrogen and
helium masses and the distribution of the $^{56}$Ni that excites the
helium.

To our knowledge there have not been any synthetic spectrum
calculations for the SN~Ib nebular phase since 1989 \cite{Fra89}.
Most of the recent effort has been devoted to inferring the
composition from emission line profiles and fluxes \cite{Mat01}.  The
oxygen mass, evidence for hydrogen, and evidence for asymmetry from
the nebular spectra of SN~1996N have been discussed by \cite{Sol98b}.

\section{SNe Ic}

\begin{figure}[ht]
\begin{center}
\includegraphics[width=.7\textwidth,angle=-90]{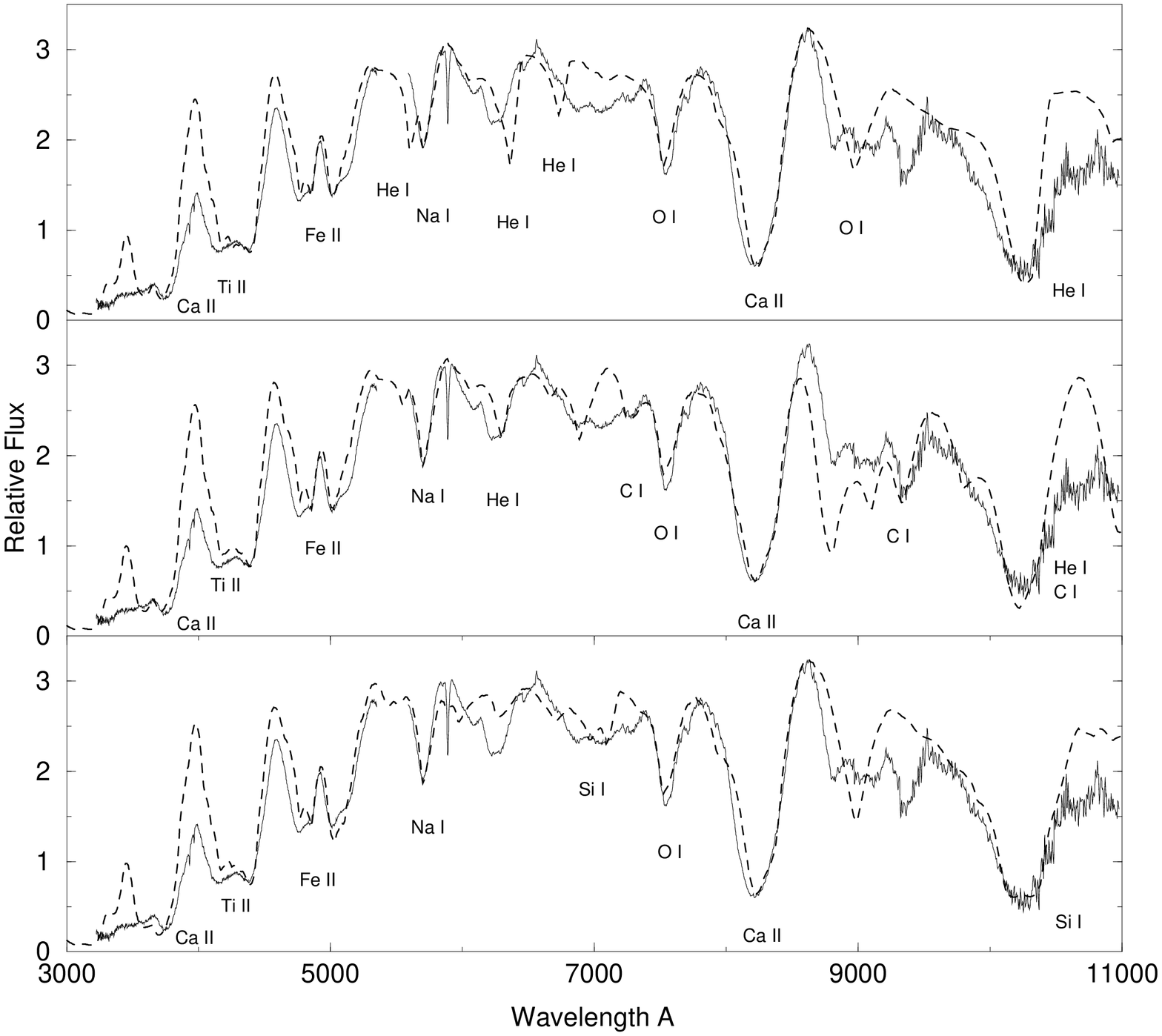}
\end{center}
\caption[]{The spectrum of the typical Type~Ic SN~1994I 7~days after
maximum brightness is compared with three \texttt{SYNOW} synthetic spectra
that have $v_{phot} = 10,000$ \kms.  To account for the observed
infrared feature, the synthetic spectra include He~I lines detached at
15,000 \kms\ ({\it top}); undetached C~I lines and He~I lines detached
at 18,000 \kms\ ({\it middle}); and Si~I lines detached at 14,000
\kms\ ({\it bottom}).  (From \cite{Mil99})\label{fig:94I}}
\end{figure}

SNe~Ic are thought to result from core collapse in massive stars that
either have lost their helium layer or fail to nonthermally excite
their helium.  If the helium layer is gone, the outer layers of the
ejected matter are expected to be mainly carbon and oxygen.  SNe~Ic
are spectroscopically more diverse than SNe~Ib \cite{Clo01,Mat01}.
The photospheric spectra are dominated by the low excitation SN
features that appear in SNe~Ib.  The spectra of the best studied
ordinary Type~Ic, SN~1994I, can be reasonably well matched by
\texttt{SYNOW} spectra \cite{Mil99} (Fig.~10).  Discrepancies between
the SN~1994I spectra and \texttt{PHOENIX} spectra indicate that the
appropriate hydrodynamical model for SN~1994I has not yet been found
\cite{Bar99}.  The important issue of whether helium is detectable in
SN~Ic spectra is not yet resolved, because an absorption near
10,000~\AA\ that sometimes is attributed to He~I $\lambda$10,830 could
be produced by one of several other ions (see Fig.~10 and
\cite{Bar99,Mil99}), and identifications of very weak optical lines
\cite{Clo97} are difficult to confirm \cite{Mat01}.

\begin{figure}[ht] \begin{center}
\includegraphics[width=.8\textwidth]{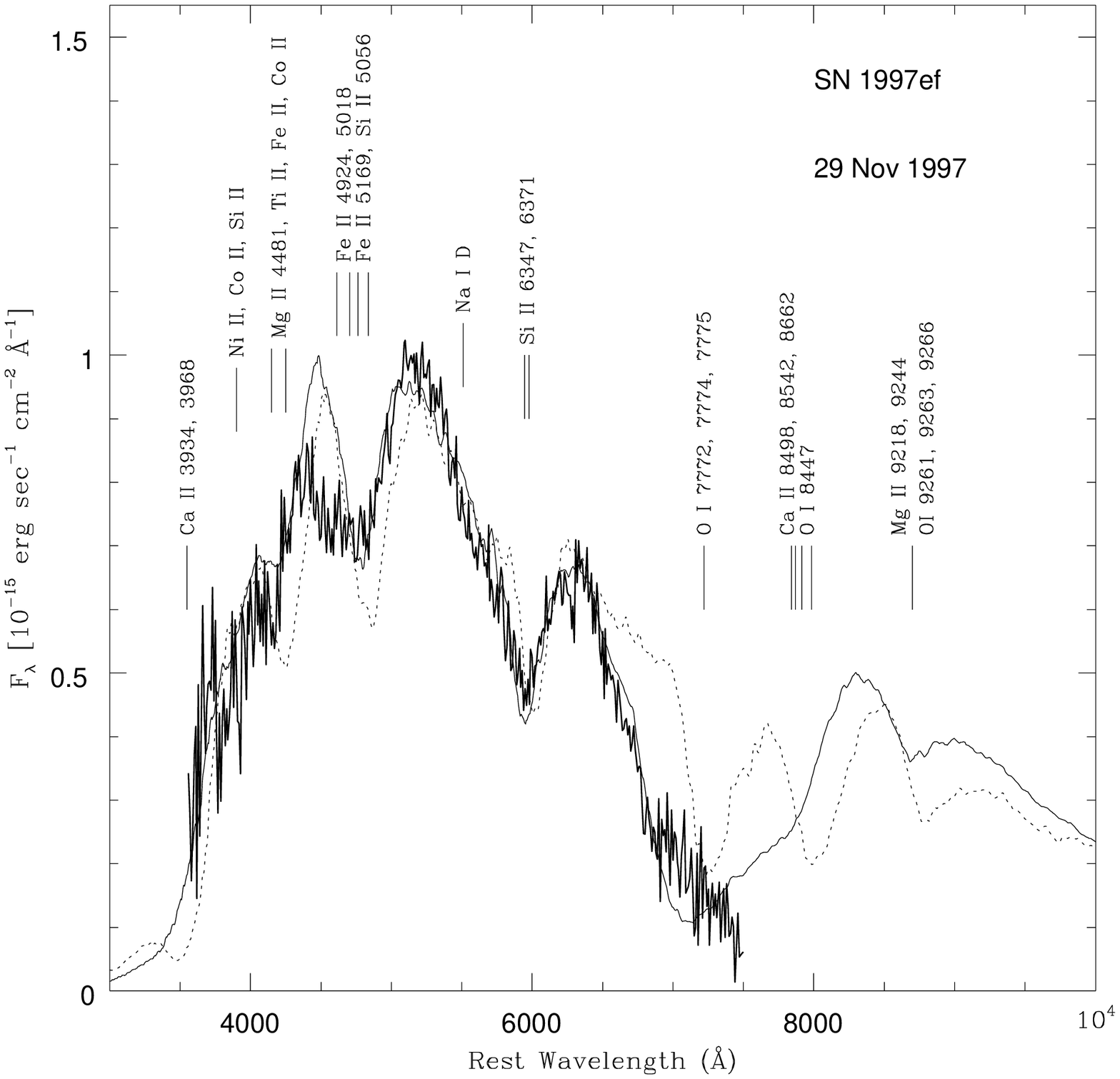} \end{center}
\caption[]{The spectrum of the hyperenergetic Type~Ic SN~1997ef about 11
days before maximum brightness({\it thick line}) is compared with two
synthetic spectra calculated with the \texttt{ML MONTE CARLO} code.
The {\it thin solid line} and the {\it dotted line} are for models
having kinetic energies of $8 \times 10^{51}$ and $1.75 \times
10^{52}$ erg, respectively.  (From \cite{Maz00b})\label{fig:97ef}}
\end{figure}

The peculiar, hyperenergetic Type~Ic SN~1998bw that was associated
with a gamma ray burst is discussed in Chap.~25.  SN~1997ef was a less
extreme example of a hyperenergetic SN~Ic.  The spectra of these
events, although difficult to interpret because of the severe Doppler
broadening and blending, can be matched fairly well by the
\texttt{SYNOW} \cite{Bra01b} and \texttt{ML MONTE CARLO} codes
\cite{Maz00b} (Fig.~11), and the resulting spectroscopic estimates of
the kinetic energy (in the spherical approximation) are far in excess
of the canonical $10^{51}$ ergs \cite{Bra01b,Maz00b,Maz01b,Pat01}.
Detailed NLTE synthetic spectrum calculations have not yet been
carried out for the photospheric spectra of hyperenergetic SNe~Ic, and
the only recent synthetic spectrum calculations for nebular phase
SNe~Ic have been for SN~1998bw with the \texttt{RL NEBULAR} code
\cite{Maz01b} (see Chap.~25).

The Type~Ic SN~1999as, one of the most luminous reliably measured SNe
to date, contained highly blueshifted ($\sim11,000$ \kms) but narrow
($\sim2000$ \kms) absorption features \cite{Hat01} that cannot be produced
by spherically symmetric SN ejecta.  Whether these unusual features
are produced by an ejecta clump in front of the photosphere or by
circumstellar matter accelerated by the SN is not yet clear.

Detailed NLTE calculations for hydrodynamical models of typical and
hyperenergetic SNe~Ic are needed, to constrain the helium mass and
the ejected mass and kinetic energy.

\section{SNe II}

SNe~II are the most spectroscopically diverse of the SN types, partly
because the optical effects of CSI range from negligible in some
events to dominant in others.  In the absence of CSI and the presence
of a substantial hydrogen envelope (SNe~II-P and II-L) the spectrum
evolves from an almost featureless continuum when the temperature is
high to one that contains first hydrogen and helium lines and then
also the usual low excitation lines as the temperature falls.  The
photospheric spectra of SN~1987A were, of course, extensively studied
\cite{Eas89,Hof88,Jef90,Sch90}. Recently it has been shown that the
observed profile of the He~I $\lambda10830$ line in SN~1987A is a
sensitive probe of the degree of mixing of $^{56}$Ni out to relatively
high velocities, $\sim3000$ \kms\ \cite{Fas99} (see also \cite{Fas98}
for similar work on the Type~II-P SN~1995V).  \texttt{PHOENIX}
calculations \cite{Mit01} show that the hydrogen lines of SN~1987A
also are a sensitive probe of the nickel mixing and may require some
$^{56}$Ni as fast as 5000 \kms.  As discussed in these three papers,
quantitative conclusions are sensitive to the details of the
macroscopic mixing of the hydrogen, helium, and nickel.

SN~1999em is a recent SN~II-P that has been extremely well observed
and that is not complicated by CSI or large departures from spherical
symmetry \cite{Leo01c}.  HST ultraviolet spectra \cite{Bar00} of this
event are discussed in Chap.~7. Two extensive independent sets of
optical spectra and photometry have been used to apply the expanding
photosphere method for determining the distance \cite{Ham01,Leo01b}.
\texttt{SYNOW} and \texttt{PHOENIX} synthetic spectra have been
compared to the photospheric spectra \cite{Bar00}.  Complicated
observed P~Cygni profiles of hydrogen and He~I lines are well
reproduced by the \texttt{PHOENIX} calculations when the density
gradient is sufficiently shallow (Fig.~12).  The total extinction can
be inferred by comparing the observed and synthetic spectra at early
times when the temperature is high \cite{Bar00}.

\begin{figure}[ht]
\begin{center}
\includegraphics[width=.7\textwidth]{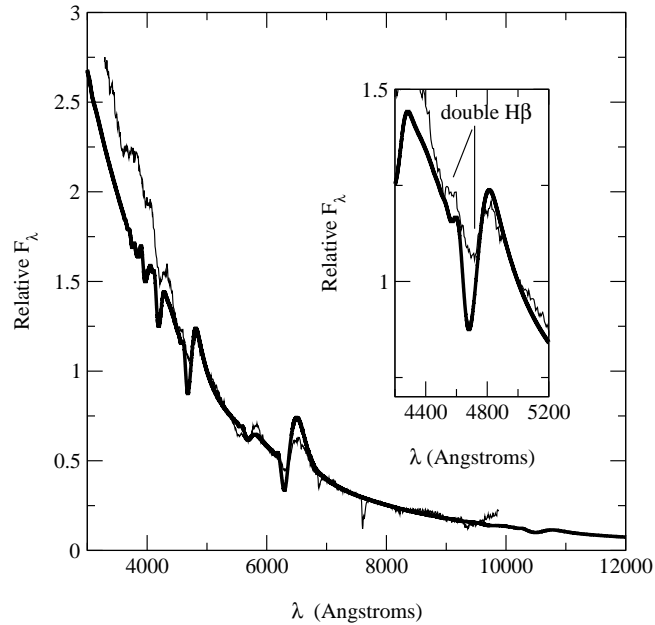}
\end{center}
\caption{The spectrum of the Type~II SN~1999em four days before maximum
brightness ({\it thin line}) is compared to a \texttt{PHOENIX} synthetic
spectrum ({\it thick line}) with $Z=Z_\odot/100$. All lines in the
synthetic spectrum are produced by hydrogen and helium. The inset
shows a shoulder in both the observed and calculated H$\beta$
profiles. (From \cite{Bar00})
\label{fig:99em}}
\end{figure}

The unusually weak Type II-P SN~1997D has attracted considerable
attention in connection with the possibility of black hole formation
and fallback.  The \texttt{ML MONTE CARLO} code has been used to fit
the spectrum with an extremely low velocity at the photosphere of only
970 \kms\ \cite{Tur98a}.  It has been pointed out that because of the
unusually low ionization, Rayleigh scattering may have significant
effects on the photospheric spectrum \cite{Chu00b}.  The low expansion
velocity offered a unique opportunity to identify lines in the nebular
spectra \cite{Ben01}.  Nebular synthetic spectra have been calculated
\cite{Chu00b} and it has been pointed out that the Sobolev
approximation may break down in the early nebular phase because
H$\alpha$ damping wings can have significant effects \cite{Chu00a}.

SNe~IIb contain conspicuous hydrogen lines at early times but not at
late times, indicating that these events lost nearly all of their
hydrogen before exploding. In the well studied SN~1993J, He~I lines
became conspicuous after maximum brightness, as the hydrogen lines faded.
The photospheric and nebular spectra were modeled in detail
(\cite{Bar95,Hof96,Hou96,Utr96} and references therein).  More
recent work on SN~1993J has concentrated on inferring properties of
the clumps and the nature of the CSI directly from line profiles and
fluxes \cite{Mat00b}.  A \texttt{SYNOW} line-identification study of
extensive observations of the Type~IIb SN~1996cb \cite{Qiu99} has been
carried out \cite{Den01}.

\begin{figure}[ht] \begin{center}
\includegraphics[width=1.0\textwidth]{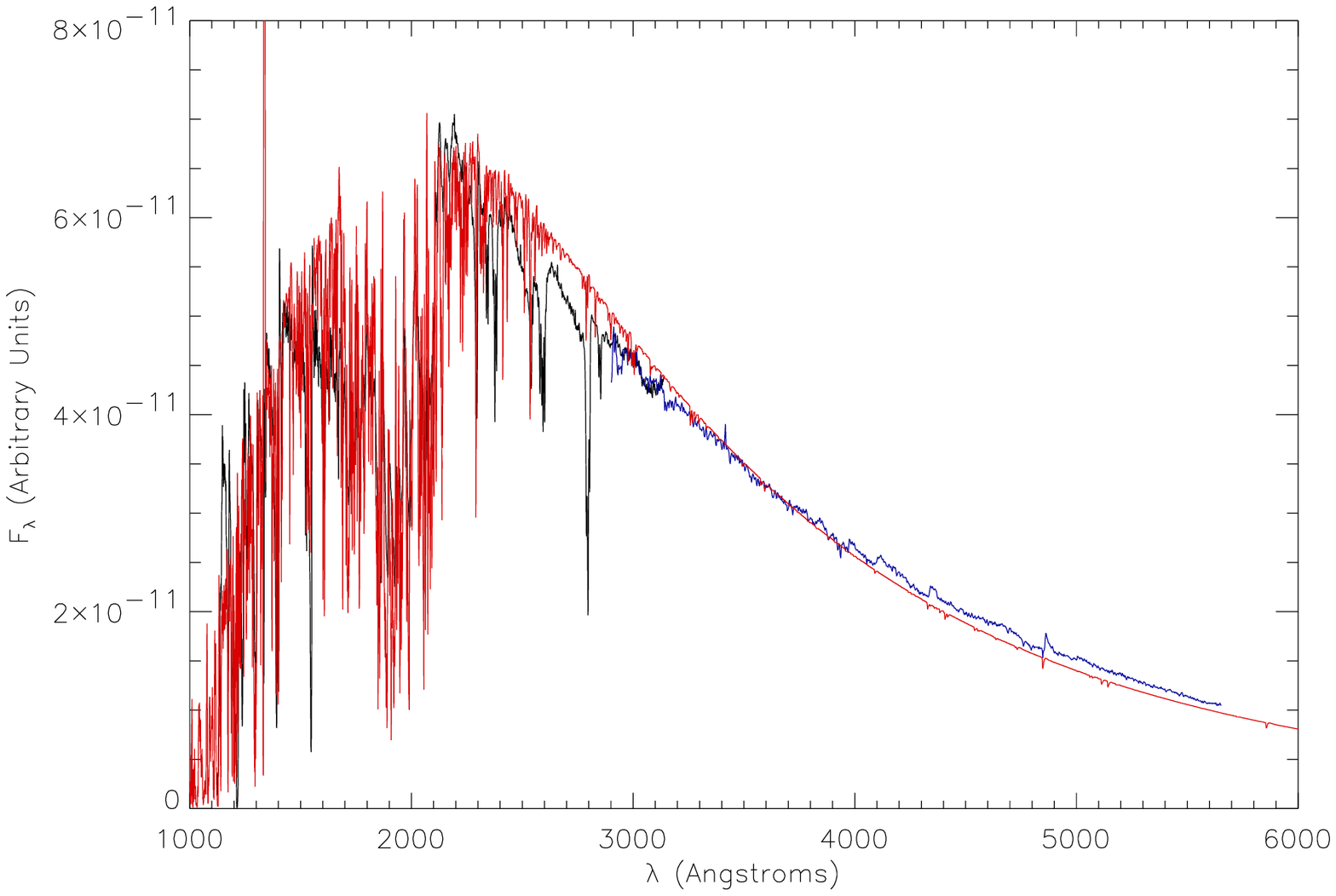} \end{center}
\caption{The UV ({\it black line}) and optical ({\it blue line})
spectra of the Type~IIn SN~1998S about four days before maximum
brightness are compared with a \texttt{PHOENIX} synthetic spectrum
({\it red line}) calculated for a circumstellar shell having an
$r^{-2}$ density distribution and a constant expansion velocity of
1000 \kms. (From Len01a)\label{fig:98s}} \end{figure}

SNe~IIn have narrow lines that form in low velocity circumstellar
matter.  Optical and infrared spectra of SN~1998S have been
intensively studied \cite{Anu01,Fas01,Len01a,Leo00a,Chu02}. HST
ultraviolet spectra \cite{Len01a} are discussed in
Chap.~7. \texttt{PHOENIX} synthetic spectra calculated for a simple
model of the circumstellar shell accounted reasonably well for the
early circumstellar-dominated spectra (Fig.~13).  The inferences from
spectroscopy are that the progenitor of SN~1998S underwent several
distinct mass loss episodes during the decades and centuries before
explosion, and that the SN ejecta or the circumstellar matter (or
both) were significantly asymmetric.  A detailed study of the nebular
spectra of the Type~IIn SN~1995N has revealed the existence of three
distinct kinematic components (Fig.~14 shows nebular phase synthetic
spectra calculated for the $\sim5000$ \kms\ intermediate--velocity
component that corresponds to unshocked SN ejecta) and produced the
suggestion that SNe~IIn are produced by red supergiants that collapse
during their superwind phases \cite{Fra01} (see also \cite{Chu97}).

\begin{figure}[ht]
\begin{center}
\includegraphics[width=.7\textwidth]{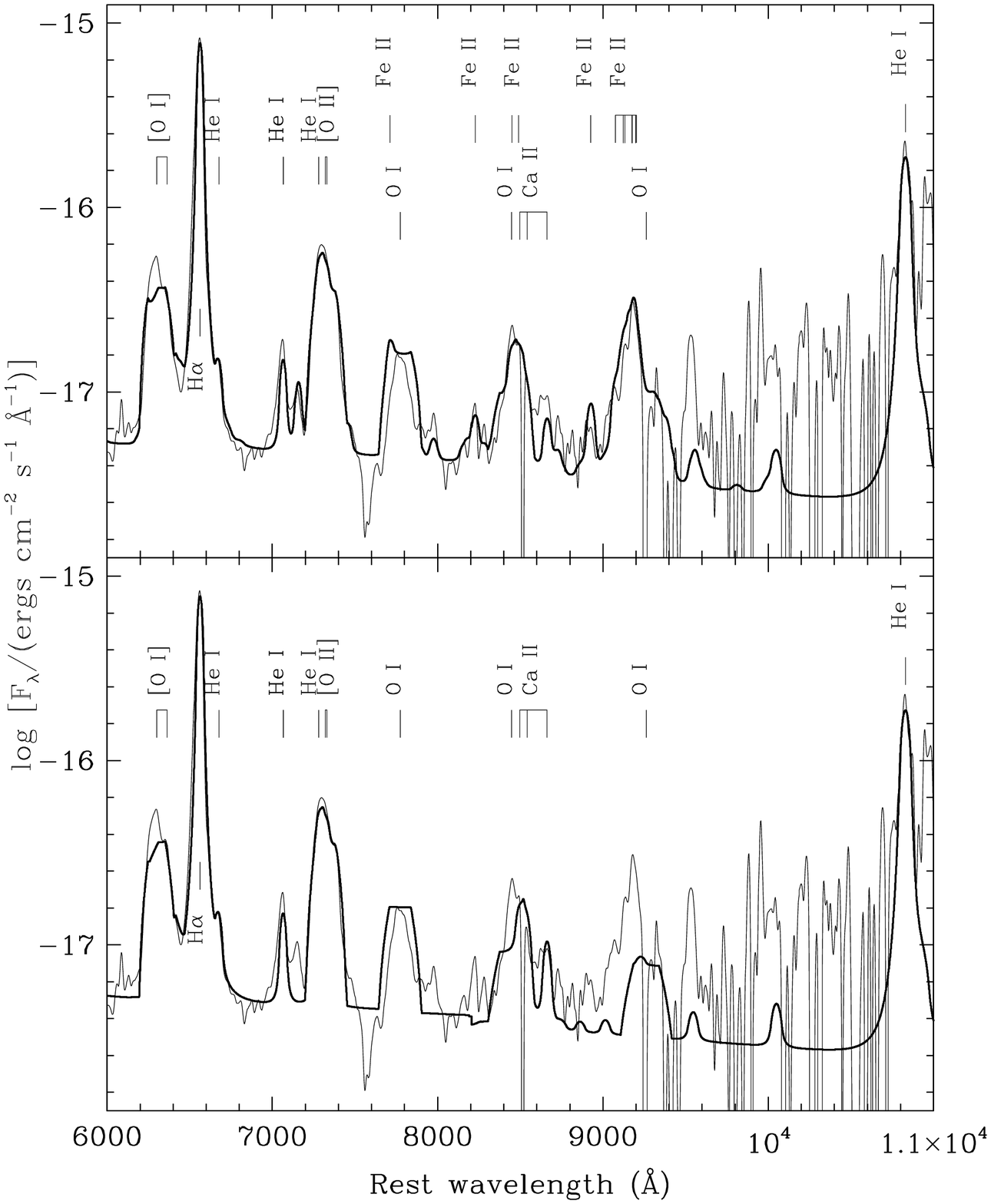}
\end{center}
\caption{The spectrum of the Type~IIn SN~1995N nearly five years after
maximum brightness is compared to synthetic spectra calculated with ({\it
upper}) and without ({\it lower}) the influence of Ly$\alpha$
fluorescence on the Fe~II lines taken into account. (From
\cite{Fra01})\label{fig:95N}}
\end{figure}

Synthetic spectra have not yet been compared to spectra of
hyperenergetic SNe~IIn such as SNe~1997cy \cite{Ger00b,Tur00} and SN~1999E
\cite{Fil00}, both of which may have been associated with gamma ray bursts.
These events may prove to be related to SN~1988Z, which has been
modeled in terms of CSI involving either dense clumps or a dense
equatorial wind \cite{Chu94}.

\section {Prospects}

Observationally, in addition to a large increase in the sheer number
of optical spectra, we can look forward to more and better infrared
and nebular spectra, spectra obtained shortly after explosion, and
spectropolarimetry.

Fast spherically symmetric (1D) photospheric--phase codes such as
\texttt{SYNOW} and \texttt{ML MONTE CARLO} will continue to be
valuable for making rapid interpretations of new flux spectra and for
certain comparative studies.  Monte Carlo 1D NLTE codes may be
constructed \cite{Luc01}. As computational resources permit, detailed
1D codes such as \texttt{PHOENIX} will be used to generate large grids
of models and spectra to expedite the comparison with observed
spectra.  \texttt{PHOENIX} will be modified to take into account the
radiative coupling between circumstellar matter and SN ejecta, and as
forbidden lines are added to the line list \texttt{PHOENIX} also will
become a powerful 1D nebular--phase code.

The major challenge for the future is to develop codes for calculating
spectra for arbitrary geometries (3D) of both the SN ejecta and the
circumstellar matter (and learning how to use such codes effectively).
For flux spectra, the already existing \texttt{CLUMPYSYN} code will be
followed in the near future by 3D Monte Carlo codes and, eventually,
by 3D \texttt{PHOENIX}--level codes.  Spectropolarization calculations
for 3D configurations are certain to emerge as a major focus of SN
spectroscopy; Monte Carlo codes for calculating 3D polarization
spectra also will become available in the near future.

We are grateful to Doug Leonard for providing an unpublished figure,
and to him and Lifan Wang for helpful comments.  This work was
supported by NASA, NSF, and the Physics Department of New Mexico Tech.


\clearpage

\begin{thebibliography}{8}
\addcontentsline{toc}{section}{References}

\bibitem{Anu01} G.C. Anupama, T. Sivarani, G. Pandey: A\&A
\textbf{367}, 506 (2001)

\bibitem{Bar95} E. Baron et~al.: ApJ \textbf{441}, 170 (1995)

\bibitem{Bar96a} E. Baron, P.H. Hauschildt, A. Mezzacappa: MN
\textbf{278}, 763 (1996)

\bibitem{Bar00} E. Baron et~al.: ApJ \textbf{545}, 444 (2000)

\bibitem{Bar99} E. Baron, D.~Branch, P.H. Hauschildt,
A.V.~Filippenko, R.P.~Kirshner: ApJ \textbf{527}, 739 (1999)

\bibitem{Bar96b} E. Baron, P.H. Hauschildt, P.~Nugent, D.~Branch:
MN \textbf{283}, 297 (1996)

\bibitem{Ben01} S. Benetti et al.: MN \textbf{322}, 361 (2001)

\bibitem{Ben98} S. Benetti, E. Cappellaro, I.J. Danziger,
M.~Turatto, F.~Patat, M.~Della Valle: MN \textbf{294}, 448 (1998)

\bibitem{Ben99} S. Benetti, M. Turatto, E. Cappellaro, I.J.~Danziger,
P.A.~Mazzali: MN \textbf{305}, 811 (1999)

\bibitem{Bow97} E.J.C. Bowers et al.: MN \textbf{290}, 663
(1997)

\bibitem{Bra01a} D. Branch: PASP \textbf{113}, 169
(2001)

\bibitem{Bra01b} D. Branch: `Direct Analysis of Spectra of Type~Ic
Supernovae'. In: \emph{Supernovae and Gamma Ray Bursts, Space
Telescope Science Institute Symposium on Supernovae and Gamma Ray
Bursts, at Baltimore, MD, USA, May~3--6, 1999}, ed. by M.~Livio,
N.~Panagia, K.~Sahu (Cambridge University, Cambridge 2001) p.~96

\bibitem{Bra02} D. Branch et al.: ApJ \textbf{566}, in press (2002)

\bibitem{Bra00} D. Branch, D.J. Jeffery, M. Blaylock, K.~Hatano:
PASP \textbf{112}, 217 (2000)

\bibitem{Cas71} J.P. Cassinelli, D.A. Hummer: MN \textbf{154}, 9
(1971)

\bibitem{Cha60} S. Chandrasekhar: {\emph Radiative Transfer} (Dover
Publications, New York 1960)

\bibitem{Che94} R.A. Chevalier, C. Fransson: ApJ \textbf{420},
268 (1994)

\bibitem{Chu92}  N.N. Chugai: 1992,
Sov. Astron. Lett. \textbf{18(3)}, 168 (1992)

\bibitem{Chu97} N.N. Chugai: Astr. Rep. \textbf{41}, 672 (1997)

\bibitem{Chu00a} N.N. Chugai: ApJ \textbf{531}, 411 (2000)

\bibitem{Chu01} N.N. Chugai: ApJ \textbf{326}, 1448 (2001)

\bibitem{Chu96} N.N. Chugai, A.A. Andronova, V.P. Utrobin:
Astr. Lett. \textbf{22}, 672 (1996)

\bibitem{Chu02} N.N. Chugai, S.I. Blinnikov, A. Fassia, P.~Lundqvist,
W.P.S.~Meikle, E.I.~Sorokina: MNRAS, in press (2002)

\bibitem{Chu94} N.N. Chugai, I.J. Danziger: MN \textbf{268}, 173
(1994)

\bibitem{Chu00b} N.N. Chugai, V.P. Utrobin: A\&A \textbf{354}, 557
(2000b)

\bibitem{Clo97} A. Clocchiatti et al: ApJ \textbf{483}, 675 (1997)

\bibitem{Clo00} A. Clocchiatti et al: ApJ \textbf{529}, 661 (2000)

\bibitem{Clo01} A. Clocchiatti et al: ApJ \textbf{553}, 886 (2001)

\bibitem{Den01} J.S. Deng, Y.L. Qiu, J.Y. Hu: ApJ, submitted (2001)

\bibitem{Den00} J.S. Deng, Y.L. Qiu, J.Y. Hu, K. Hatano, D.~Branch:
ApJ \textbf{540}, 452 (2000)

\bibitem{Eas89} R. Eastman, R.P. Kirshner: ApJ \textbf{347}, 771
(1989)

\bibitem{Fas01} A. Fassia et al.: MN \textbf{325}, 907 (2001)

\bibitem{Fas99} A. Fassia, W.P.S. Meikle: MN \textbf{302}, 314
(1999)

\bibitem{Fas98} A. Fassia, W.P.S. Meikle, T.R. Geballe,
N.A.~Walton, D.L.~Pollacco, R.G.M.~Rutten, C.~Tinney: MN
\textbf{299}, 150 (1998)

\bibitem{Fes99} R. Fesen et al.: AJ \textbf{117}, 725 (1999)

\bibitem{Fil97} A.V. Filippenko: ARAA \textbf{35}, 309 (1997)

\bibitem{Fil00} A.V. Filippenko: `Optical Observations of Type~II
Supernovae'. In: \emph{Cosmic Explosions, 10th Annual October
Astrophysics Conference, at College Park, MD, October~11-13, 1999},
ed. by S.~Holt, W.W.~Zhang (AIP, New York 2000)

\bibitem{Fis00} A.K. Fisher: Direct Analysis of Type Ia Supernova
Spectra. PhD Thesis, University of Oklahoma, Norman (2000)

\bibitem{Fis99} A.K. Fisher, D. Branch, K. Hatano, E. Baron:
MN \textbf{304}, 67 (1999)

\bibitem{Fra84} C. Fransson: A\&A \textbf{132}, 115  (1984)

\bibitem{Fra94} C. Fransson: `The Late Emission from
Supernovae'. In: \emph{Supernovae, NATO Advanced Study Institute on
Supernovae, at Les Houches, France, July~31--September~1, 1990},
ed. by S.A.~Bludman, R.~Mochkovitch, J.~Zinn-Justin (Elsevier,
Amsterdam 1994) p.~677

\bibitem{Fra01} C. Fransson et al.: ApJ, in press (2001)

\bibitem{Fra89} C. Fransson, R.A. Chevalier: ApJ \textbf{343}, 323
(1989)

\bibitem{Gar01} P.M. Garnavich et al.: ApJ, in press (2001)

\bibitem{Ger00a} C.L. Gerardy, R.A. Fesen, P. H\"oflich,
J.C.~Wheeler: AJ \textbf{119}, 2968, (2000)

\bibitem{Ger00b} L.M. Germany, D.J. Riess, E.M. Sadler,
B.P. Schmidt, C.W. Stubbs: ApJ \textbf{533}, 320, (2000)

\bibitem{Ham01} M. Hamuy et al.: ApJ \textbf{558}, 615 (2001)

\bibitem{Har87} R.P. Harkness et al.: ApJ \textbf{317}, 355 (1987)


\bibitem{Hat99a} K. Hatano, D. Branch, A. Fisher, E. Baron,
A.~V.~Filippenko: ApJ \textbf{525}, 881 (1999)

\bibitem{Hat99b} K. Hatano, D. Branch, A. Fisher, J. Millard,
E.~Baron: ApJ Suppl. \textbf{121}, 233 (1999)

\bibitem{Hat00} K. Hatano, D. Branch, E.J. Lentz, E.~Baron,
A.V. Filippenko, P.M. Garnavich: ApJ \textbf{543}, L49 (2000)

\bibitem{Hat01} K. Hatano, D. Branch, K. Nomoto, J.S. Deng,
K.~Maeda, P.~Nugent, G.~Aldering: BAAS \textbf{198}, 3902 (2001)

\bibitem{Hau99} P.H. Hauschildt, E. Baron: J.~Comp. Appl. Math.
\textbf{109}, 41 (1999)

\bibitem{Her00} M. Hernandez et al.: MN \textbf{319}, 223 (2000)

\bibitem{Hof88} P. H\"oflich : Proc. Astron. Soc. Aust. \textbf{7},
434 (1988)

\bibitem{Hof91} P. H\"oflich: A\&A \textbf{246}, 481 (1991)

\bibitem{Hof01} P. H\"oflich, A. Khokhlov, L. Wang: `Aspherical
Supernova Explosions: Hydrodynamics, Radiation Transport, \&
Observational Consequences'.  In: \emph{Relativistic Astrophysics,
20th Texas Symposium on Relativistic Astrophysics, at Austin, TX,
December~10-15, 2000}, ed. by J.C.~Wheeler, H.~Martel (New York, AIP,
2001) p.~459

\bibitem{Hof96} P. H\"oflich, J.C. Wheeler, D.C. Hines,
S.R. Tramell: ApJ \textbf{459}, 307 (1996)

\bibitem{Hof98} P. H\"oflich, J.C. Wheeler, F.K. Thielemann:
ApJ \textbf{495}, 617 (1998)

\bibitem{Hou96} J.C. Houck, C. Fransson: ApJ \textbf{456}, 811
(1996)

\bibitem{How01} D.A. Howell, P. H\"oflich, L. Wang, J.C. Wheeler:
ApJ \textbf{556}, 302 (2001)

\bibitem{Jef89} D.J. Jeffery: ApJS \textbf{71}, 951 (1989)

\bibitem{Jef91} D.J. Jeffery: ApJS \textbf{77}, 405 (1991)

\bibitem{Jef93} D.J. Jeffery: ApJ \textbf{415}, 734 (1993)

\bibitem{Jef01} D.J. Jeffery: ApJ, submitted (2001)

\bibitem{Jef90} D.J. Jeffery, D. Branch: `Analysis of Supernova
Spectra'. In: \emph{Supernovae, 6th Jerusalem Winter School for
Theoretical Physics, at Jerusalem, Israel, December~28, 1988--January
5, 1989}, ed. by J.C.~Wheeler, T.~Piran, S.~Weinberg (World
Scientific, Singapore 1990) p.~149

\bibitem{Jha99} S. Jha et~al.: ApJ Suppl. \textbf{125}, 73 (1999)

\bibitem{Karp77} A.H. Karp, G. Lasher, K.L. Chan, E.E. Salpeter:
ApJ \textbf{214}, 161 (1977)

\bibitem{Kas01} D. Kasen, D. Branch, E. Baron, D.J. Jeffery:
ApJ, in press (2001)

\bibitem{Len01a} E.J. Lentz et~al.: ApJ \textbf{547}, 406 (2001)

\bibitem{Len01b} E.J. Lentz, E. Baron, D. Branch, P. Hauschildt:
ApJ \textbf{547}, 402 (2001)

\bibitem{Len01c} E.J. Lentz, E. Baron, D. Branch, P. Hauschildt:
ApJ \textbf{557}, 266 (2001)

\bibitem{Len00} E.J. Lentz, E. Baron, D. Branch, P. Hauschildt,
P.E.~Nugent: ApJ \textbf{530}, 966 (2000)

\bibitem{Leo01a} D.C. Leonard et al.: PASP, in press (2001)

\bibitem{Leo01b} D.C. Leonard, A.V. Filippenko: PASP \textbf{113},
920 (2001)

\bibitem{Leo01c} D.C. Leonard, A.V. Filippenko, D.R. Ardila,
M.S. Brotherton: ApJ \textbf{553}, 861 (2001)

\bibitem{Leo00a} D.C. Leonard, A.V. Filippenko, A.J. Barth,
T. Matheson: ApJ \textbf{536}, 239 (2000)

\bibitem{Leo00b} D.C. Leonard, A.V. Filippenko, T. Matheson: `Probing
the Geometry of Supernovae with Spectropolarimetry'.  In: \emph
{Cosmic Explosions, 10th Annual October Astrophysics Conference at
College Park, MD, October~11-13, 1999}, ed. by S.~Holt, W.~W.~Zhang
(AIP, New York 2000), p.~165

\bibitem{Li93} H. Li, R. McCray, R.A. Sunyaev: ApJ \textbf{419},
824 (1993)

\bibitem{Li99} W.D. Li et al.: ApJ
\textbf{117}, 2709 (1999)

\bibitem{Li01} W.D. Li et al.: PASP \textbf{113}, 1178 (2001)

\bibitem{Li00} W.D. Li, A.V. Filippenko, R.R. Treffers,
A.G. Riess, J.~Hu, Y.~Qiu: ApJ \textbf{546}, 734 (2000)

\bibitem{Liu98} W. Liu, D.J. Jeffery, D.R.Schultz: ApJ
\textbf{494}, 812 (1998)

\bibitem{Luc91} L.B. Lucy: ApJ \textbf{383}, 308 (1991)

\bibitem{Luc99} L.B. Lucy: A\&A \textbf{345}, 211 (1999)

\bibitem{Luc01} L.B. Lucy: MN \textbf{326}, 95 (2001)

\bibitem{Mat00a} T. Matheson, A.V. Filippenko, R. Chornock,
D.C.~Leonard, W.~Li: AJ \textbf{119}, 2303 (2000)

\bibitem{Mat00b} T. Matheson, A.V. Filippenko, Ho, L.C., Barth,
A.J., D.C.~Leonard: AJ \textbf{120}, 1499 (2000)

\bibitem{Mat01} T. Matheson, A.V. Filippenko, W. Li, D.C.~Leonard,
J.C.~Shields: AJ \textbf{121}, 1648 (2001)

\bibitem{Maz98a} P.A. Mazzali: ApJ \textbf{499}, L49 (1998)

\bibitem{Maz00a} P.A. Mazzali: A\&A \textbf{363}, 705 (2000)

\bibitem{Maz01a} P.A. Mazzali: MN \textbf{321}, 341 (2001)

\bibitem{Maz97} P.A. Mazzali, N. Chugai, M. Turatto, L.B.~Lucy,
I.J.~Danziger, E.~Cappellaro, M.~Della Valle, S.~Benetti: MN
\textbf{284}, 151 (1997)

\bibitem{Maz00b} P.A. Mazzali, K. Iwamoto, K. Nomoto: ApJ \textbf{545}, 407 (2000)

\bibitem{Maz98b} P.A. Mazzali, L.B. Lucy: MN \textbf{295}, 428
(1998)

\bibitem{Maz01b} P.A. Mazzali, K. Nomoto, F. Patat, K.~Maeda:
ApJ 559, 1047 (2001)

\bibitem{Mcc84} M.L. McCall: MN \textbf{210}, 829 (1984)

\bibitem{Mei96} W.P.S. Meikle et al.: MN \textbf{281}, 263 (1996))

\bibitem{Mih78} D. Mihalas: {\emph Stellar Atmospheres,}
(W. H. Freeman, San Francisco 1978)

\bibitem{Mil99} J. Millard et al.: ApJ \textbf{527}, 746 (1999)

\bibitem{Mil01} P.A. Milne, L.-S. The, M.D. Leising: ApJ
\textbf{559}, 1019 (2001)

\bibitem{Mit01} R. Mitchell, E. Baron, D. Branch, P. Lundqvist,
S. Blinnikov, P.H.~Hauschildt, C.S.J. Pun: ApJ \textbf{556}, 979
(2001)

\bibitem{Mod01} M. Modjaz, W. Li, A.V. Filippenko, J.Y. King,
D.C. Leonard, T.~Matheson, R.R.~Treffers: PASP \textbf{113}, 308
(2001)

\bibitem{Nom84} K. Nomoto, F.-K. Thielemann, Y. Yokoi: ApJ
\textbf{286}, 644 (1984)

\bibitem{Nug95} P. Nugent, M.M. Phillips, E. Baron, D. Branch,
P.~Hauschildt: ApJ \textbf{455}, L147 (1995)

\bibitem{Pat01} F. Patat et al.:
ApJ \textbf{555}, 900 (2001)

\bibitem{Pin00} P. Pinto, R.G. Eastman:: ApJ \textbf{530}, 757 (2000)

\bibitem{Qiu99} Y. Qiu, W. Li, Q. Qiao, J. Hu: AJ \textbf{117}, 736
(1999)

\bibitem{Rui95} P. Ruiz-Lapuente: ApJ \textbf{439}, 60 (1995)


\bibitem{Rui92} P. Ruiz-Lapuente, L.B. Lucy: ApJ \textbf{400}, 127
(1992)

\bibitem{Sal01} M.E. Salvo, E. Cappellaro, P.A. Mazzali,
S.~Benetti, I.~J.~Danziger, M.~Turatto: MN \textbf{321}, 254
(2001)

\bibitem{Sal98} I. Salamanca, R. Cid-Fernandes, G. Tenorio-Tagle,
E.~Telles, R.J.~Terlevich, C.~Munoz-Tunon: MN \textbf{300}, 17
(1998)

\bibitem{Sch90} W. Schmutz, D.C. Abbott, R.S. Russell,
W.-R. Hamann, U.~Wessolowski: ApJ \textbf{355}, 255 (1990)

\bibitem{Ser73} K. Serkowski: In \emph { Interstellar Dust and
Related Topics, IAU Symposium 52, at Albany, NY, May~29 -- June~2,
1972} ed. by J.~M.~Greenberg, H.~C.~van~de~Hulst (Reidel, Dordrecht
1973), p.~145

\bibitem{Sha82} P.R. Shapiro, P.G. Sutherland: ApJ \textbf{263},
902 (1982)

\bibitem{Sol98a} J. Sollerman, R.J. Cumming, P. Lundqvist: A\&A
\textbf{493}, 933 (1998)

\bibitem{Sol98b} J. Sollerman, B. Leibundgut, J. Spyromilio: A\&A
\textbf{337}, 207 (1998)

\bibitem{Spy01} J. Spyromilio, B. Leibundgut, R. Gilmozzi: A\&A
\textbf{376}, 188 (2001)

\bibitem{Swa93} D.A. Swartz, A.V. Filippenko, K. Nomoto,
J.C. Wheeler: ApJ \textbf{411}, 313 (1993)

\bibitem{Swa95} D.A. Swartz, P.G. Sutherland, R.P. Harkness
J.C. Wheeler: ApJ \textbf{446}, 766 (1995)

\bibitem{Thi86} F.-K. Thielemann, K. Nomoto, K. Yokoi:  A\&A,
                   \textbf{158}, 17 (1986)

\bibitem{Tho02} R.C. Thomas, D. Kasen, D. Branch, E. Baron: ApJ
\textbf{567}, in press (2002)

\bibitem{Tur98a} M. Turatto et~al.: ApJ \textbf{498}, L129 (1998)

\bibitem{Tur00} M. Turatto et~al.: ApJ \textbf{534}, L57 (2000)

\bibitem{Tur98b} M. Turatto, A. Piemonta, S. Benetti, E. Cappellaro,
P.A. Mazzali, I.J. Danziger, F. Patat: AJ \textbf{116}, 2431 (1998)

\bibitem{Utr96} V.P. Utrobin: A\&A \textbf{306}, 231 (1996)

\bibitem{Vin01} J. Vink\'o, L.L. Kiss, B. Cs\'ak, G.~F\"ur\'esz,
R.~Sxab\'o, J.R.~Thomson, S.W.~Mochnacki:  \textbf{121}, 3127 (2001)

\bibitem{Wan96} L. Wang, J.C. Wheeler: ApJ \textbf{462}, L27 (1996)

\bibitem{Wan97} L. Wang, J.C. Wheeler, P. H\"oflich: ApJ
\textbf{476}, L27 (1997)

\bibitem{Wan01a} L. Wang, D.A. Howell, P. H\"oflich, J.C. Wheeler:
ApJ \textbf{550}, 1030 (2001)

\bibitem{Wan01b} L. Wang et al.: Nature, submitted (2001)

\bibitem{Whe00} J.C. Wheeler, S. Benetti: `Supernovae'. In: {\emph
Allen's Astrophysical Quantities,} 4th Edition, ed. by A.~N.~Cox
(Springer, New York 2000), p.~451

\bibitem{Whe98} J.C. Wheeler, P. H\"oflich, R.P. Harkness,
J.~Spyromilio: ApJ \textbf{496}, 908 (1998)

\bibitem{Whe93} J.C. Wheeler, D.A. Swartz, R.P. Harkness:
Phys.~Rep. \textbf{227}, 113 (1993)

\bibitem{Woo97} S.E. Woosley, R.G Eastman: `Type~Ib and Ic
Supernovae: Models and Spectra'. In: \emph{Thermonuclear Supernovae,
NATO Advanced Study Institute on Thermonuclear Supernovae, at
Aiguablava, Spain, June~20--30, 1995}, ed. by P. Ruiz-Lapuente,
R. Canal, J. Isern (Kluwer, Dordrecht 1997) p.~821



\end{thebibliography}
\end{document}